\documentclass[a4paper]{article}

\usepackage[margin=1in]{geometry} 

\usepackage{amsmath}
\usepackage{amsthm}
\usepackage{amssymb}
\usepackage{esint}
\usepackage{graphicx}
\usepackage{subcaption}
\usepackage{float}
\usepackage{xcolor}

\allowdisplaybreaks
\usepackage[utf8]{inputenc}
\usepackage[hidelinks]{hyperref}
\hypersetup{
	unicode,
	pdfauthor={Author One, Author Two, Author Three},
	pdftitle={OPY2025},
	pdfsubject={A simple article template},
	pdfkeywords={article, template, simple},
	pdfproducer={LaTeX},
	pdfcreator={pdflatex}
}

\newcommand{\C}{\mathbb{C}}

\newcommand{\Z}{\mathbb{Z}}

\newcommand{\R}{\mathbb{R}}

\newcommand{\sech}{\mathrm{sech}}

\newtheorem{thm}{Theorem}[section]
\newtheorem{lemma}[thm]{Lemma}

\theoremstyle{remark}

\newcommand{\bd}{\boldsymbol}

\newcommand{\beq}{\begin{equation}}	
	\newcommand{\eeq}{\end{equation}}
\newcommand{\beqna}{\begin{eqnarray*}}
	\newcommand{\eeqna}{\end{eqnarray*}}
\newcommand{\beqn}{\begin{equation*}}	
	\newcommand{\eeqn}{\end{equation*}}
\newcommand{\bp}{\begin{proof}}	
	\newcommand{\ep}{\end{proof}}
\newcommand{\bprop}{\begin{proposition}}
	\newcommand{\eprop}{\end{proposition}}
\newcommand{\bt}{\begin{theorem}}	
	\newcommand{\et}{\end{theorem}}
\newcommand{\bc}{\begin{corollary}}
	\newcommand{\ec}{\end{corollary}}
\newcommand{\bl}{\begin{lemma}}	
	\newcommand{\el}{\end{lemma}}

\title{Soliton and breather interactions in the integrable discrete focusing Manakov system via  Hirota's method}
\author{Uyen Le$^1$,\;\; Alexander Chernyavsky$^{2,3}$ and Barbara Prinari$^{4,5,*}$}
\date{
    \small{$^1$ Department of Mathematics \& Statistics, Texas A\&M University Corpus Christi, TX 78412, USA\\
    $^2$ Department of Mathematics \& Statistics, Brock University, St Catharines, L2S 3A1
 ON, Canada \\
 {$^3$ Akian College of Science \& Engineering, American University of Armenia, Yerevan, Armenia}\\
    $^{4}$ Department of Mathematics, State University of New York, Buffalo, NY 14260, USA\\
    $^5$ Department of Mathematics, University of Ioannina, Ioannina 45110, Epirus, Greece}\\
    $^*$ \href{mailto:bprinari@buffalo.edu}{Corresponding author: bprinari@buffalo.edu}
}

\begin{document}

\maketitle

\begin{abstract}
The aim of this paper is to apply Hirota’s bilinear method to the integrable discrete Manakov system in the focusing dispersion regime in order to construct and analyze soliton and breather solutions. After deriving the general bilinear form of the system, we show how to obtain fundamental solitons, as well as fundamental and composite breathers. We then obtain solutions exhibiting 2 solitons and 2 breathers and combinations of a soliton and a breather, and discuss all ``two-body'' interactions properties, with particular emphasis on explicit formulas, visualization, and long-time asymptotic behavior, thus rigorously confirming the highly nontrivial interaction properties of these coherent structures. 
\end{abstract}

\section{Introduction}
Integrable nonlinear wave equations play a central role in applied mathematics and mathematical physics, as they usually represent analytically tractable models exhibiting nonlinear phenomena such as solitons, breathers, and other coherent structures. Since the development of the Inverse Scattering Transform (IST) almost 60 years ago \cite{GardnerGreeneKruskalMiura1967,ZS72,AblowitzSegur1981}, integrable systems have provided a fundamental framework for the analysis of nonlinear wave propagation. In particular, the celebrated nonlinear Schr\"odinger (NLS) equation \cite{ZS72}, and a coupled NLS equation --the so-called Manakov system-- introduced by Manakov \cite{[1]} in 1974, have been widely studied over the years also due to their relevance in nonlinear optics, Bose–Einstein condensates, and scalar and multicomponent wave dynamics \cite{KivsharAgrawal2003,[2],PitaevskiiStringari2016}.

Alongside continuous models, discrete integrable systems have received sustained attention for over 50 years. Besides {being} crucial for numerical schemes, discretizations arise naturally in applications such as optical waveguide arrays and lattice dynamical systems, and preserving integrability ensures the retention of key qualitative features, including elastic interactions and infinite sequences of conserved quantities. A prototypical example is the Ablowitz–Ladik (AL) lattice \cite{[3],[3']}, whose multicomponent generalization, namely
the following system of differential-difference equations:
	\begin{equation}\label{manakov}
		i \frac{d \bd q_n}{dt} = \frac{1}{{\Delta h^2}} \left( \bd q_{n+1} - 2 \bd q_n + \bd q_{n-1} \right) - \sigma \|\bd q_n \|^2 \left(\bd q_{n+1} +\bd q_{n-1} \right), \qquad \sigma = \mp 1,
	\end{equation}
where $\bd q_n(t)$ is a 2-component complex vector function of $n\in \Z$, $t\in \R$, while $h\in \mathbb{R}$ represents the lattice spacing, and $\sigma = \mp 1$ distinguishes between the focusing/defocusing dispersion regimes,  provides an $O(h^2)$ integrable discrete analog of the Manakov system, to which it reduces in the limit as ${\Delta h}\to 0, n{\Delta h}\to x$. This model was introduced as a vector generalization of the AL equations \cite{GI81a,GI81b,GI82,TUW99,AOT99,Tsuchida00}, and is an integrable spatial discretization of the continuous Manakov system, which we refer to as integrable discrete Manakov (IDM) system. In this work, we consider only the focusing regime and take $\sigma = -1$ throughout. One could also take ${\Delta h}=1$ without loss of generality, since the parameter can be removed by a simple rescaling of the dependent field.

The inverse scattering theory, presented, for instance, in a unified framework for discrete and continuous NLS systems in \cite{[4]}, completely linearizes the associated initial-value problem, and in addition it provides unique insight into the spectral properties of explicit soliton solutions, which can be constructed as a by-product of the IST. For the IDM system \eqref{manakov}, the IST framework developed in \cite{[4],APT04b,APT06} showed that the spectrum consists of symmetric octets of discrete eigenvalues, with associated $2\times 2$ norming constants. The nature of the soliton crucially depends on the rank of the associated norming constants. Specifically, when the rank of the norming constants is 1, for a single octet of discrete eigenvalues one obtains either a {fundamental} soliton (FS), when one of the columns (or rows) of the norming constants is identically zero, or a {fundamental} breather (FB), if the columns (or rows) are proportional. A FS is the natural discretization of a bright soliton of the Manakov system, while a FB is an orthogonal superposition of 2 fundamental solitons, with the same amplitudes and velocities and opposite carriers. When the norming constants are of full rank, the corresponding solutions are referred to as composite breathers (CB), and are more complicated superpositions of fundamental solitons, still with the same amplitude and traveling with the same velocity. It should be stressed that fundamental and composite breathers are purely discrete solutions, and do not have a continuous counterpart in the Manakov system.

On the other hand, exact solutions of integrable systems can be typically constructed more efficiently through several complementary direct methods, e.g., Hirota's bilinear formalism, Darboux transformations, and algebro-geometric techniques. Among these, Hirota’s approach is particularly effective because of its direct and constructive nature. By reformulating nonlinear equations into bilinear form via suitable dependent variable transformations, the method enables the systematic construction of multi-soliton solutions through perturbative expansions \cite{[7],[8],[9]}. Moreover, it extends naturally to differential–difference equations and fully discrete systems, where it provides a unified framework for generating soliton, breather, and higher-order solutions.

In  \cite{Ohta00,[5]}, Pfaffians were used to obtain a bilinear formulation and explicit multi-soliton solutions for a discrete vector system equivalent to the one considered in this work. However, those works are primarily concerned with fundamental soliton solutions, and do not address the construction of breather-type excitations. In particular, neither fundamental breathers nor composite breathers are presented in \cite{Ohta00,[5]}, leaving open the problem of systematically constructing and classifying such oscillatory structures within the Hirota's formalism. {
In \cite{DvdM11,DvdM12}  explicit solutions of the matrix integrable discrete nonlinear Schr\"odinger equation were obtained in terms of matrix triplets, by using the IST and solving the Marchenko equation of the inverse problem by separation of variables (see also \cite{DvdM13} for an alternative formulation of the IST for integrable discrete
nonlinear Schr\"odinger equations). We point out, however, that the explicit solution provided in Example 6.2 in \cite{DvdM12} is a fundamental soliton. One can expect that fundamental and composite breathers could also be obtained by solving appropriate linear systems from the Marchenko equations, but, again, identifying and characterizing the necessary constraints on the matrix triplets that lead to the breather excitations is currently an open problem, to the best of our knowledge. 
}
To further situate the above classes of solutions within the existing literature, breather interactions in the IDM system were recently studied in \cite{13}, where the explicit solutions are expressed in terms of ratios of determinants, yielding compact representations which however can become unwieldy in practice. In particular, for multi-soliton and multi-breather configurations, the size of the determinants grows rapidly with the number of interacting modes. Even in relatively simple cases—such as two-soliton solutions or mixed configurations involving one soliton and one breather—the explicit evaluation and visualization of these formulas can be technically demanding.
This difficulty has significant implications. First, plotting such solutions typically requires substantial symbolic or numerical effort, which can obscure their qualitative features. Second, the computation of long-time asymptotic behavior is nontrivial, as it involves delicate asymptotic analysis of determinant expressions and the identification of dominant contributions. As a result, although \cite{13} provides a detailed study of breather interactions and reveals novel dynamical properties of the discrete breathers, the direct use of the solution formulas for visualization and asymptotic characterization remains comparatively intricate.

By contrast, Hirota’s bilinear method typically yields solution expressions that are immediately amenable to both analytical and numerical treatment. Multi-soliton and multi-breather solutions are expressed as finite sums of exponential terms, making the identification of dominant contributions in different asymptotic regimes straightforward. This facilitates the computation of long-time asymptotics, including phase shifts and interaction-induced effects. Furthermore, the explicit form of the solutions enables efficient numerical evaluation and direct visualization, even for configurations involving multiple interacting structures. In particular, interactions among fundamental solitons, fundamental breathers, and composite breathers can be analyzed and illustrated in a transparent manner.

The aim of this work is to apply Hirota’s bilinear method to the IDM system \eqref{manakov} in order to construct and analyze fundamental soliton and breather solutions, as well as their composite counterparts. 
The paper is organized as follows. In Section 2, we derive the bilinear form, and use it to obtain the basic solutions: fundamental soliton, fundamental breather, and composite breather. Although these solutions are already known, to the best of the authors' knowledge the derivation of the breathers via Hirota's method is new. Section 3 presents the construction of solutions with pairs of all possible types of solitons and breathers, and investigates their interaction properties by explicitly computing the long-time dynamics. As mentioned above, these interaction properties were studied in \cite{13}, but the solutions were written as ratios of determinants of $8\times 8$ matrices, which made {plotting} them quite involved. Moreover, their long-time asymptotics in the limit in which the exponentials grow was not computed, as the matrices involved in the computation become degenerate to leading order, and one or two next-to-leading order terms are necessary to resolve the indeterminacy. In \cite{13}, this complication was circumvented by applying a heuristic method (the Manakov method, see \cite{[1]}) to gain insight on the long-time asymptotics. Although useful, the Manakov method has an intrinsic limitation in that it requires certain a priori assumptions on the form of the solution in the long-time limit, while the present work allows to directly compute the asymptotics in both time directions. Conclusions and perspectives are given in Section 4.

\section{Explicit solutions via Hirota's bilinear method} 
We outline here the three main steps in Hirota's bilinear method. First, we seek to replace the IDM system by an equation (or system of equations) which is homogeneous in degree. Similarly to how this is achieved in the scalar NLS equation, we consider the following ansatz for the dependent variable:
	\begin{equation}\label{hirota var}
		\bd q_n(t) = \frac{1}{f_n}\left(\begin{array}{c}
			g_n \\ h_n
		\end{array}\right)
	\end{equation}
where $g_n$ and $h_n$ are complex-valued function of $n\in \Z$ and $t\in \R$, whereas $f_n$ is a real valued function over the same domain. Next, substituting the transformation \eqref{hirota var} in the IDM equation \eqref{manakov} and decoupling the resulting equations appropriately so that they are homogeneous in degree which is quadratic in this case, we get
	\begin{subequations}\label{bilinear}
		\begin{align}
			i \Delta h^2 D_t g_n f_n  &= f_{n-1} g_{n+1} - 2f_n g_{n} + f_{n+1} g_{n-1},  \\
			i \Delta h^2 D_t h_n f_n  &= f_{n-1} h_{n+1} - 2f_n h_{n} + f_{n+1} h_{n-1}, \\
			f_{n+1}f_{n-1} - f_{n}^2 &=  \Delta h^2\left(g_n g^{*}_n+ h_n h^{*}_n\right),
		\end{align}
	\end{subequations}
where $*$ denotes the complex conjugation, and the Hirota's differential operator $D_t$ is defined by 
	\begin{equation}\label{hirota D op}
		D^{m}_t (u\cdot v) = (\partial_t - \partial_{t'})^m u(n,t)v(n,t')\Big\vert_{t'=t}
	\end{equation}
with $m\in \Z^{+} $. We observe that the Hirota's operator acts similarly to the usual derivative product rule but with alternating signs:
	\begin{equation}
		D^{m}_t (u\cdot v) = \sum_{k=0}^{m}\frac{(-1)^{m-k}m!}{k!(m-k)!}\left(\frac{\partial^{k}u}{\partial t^{k}}\right)\left(\frac{\partial^{m-k}v}{\partial t^{m-k}}\right).
	\end{equation}
Finally, we seek solutions for $g_n$, $h_n$ and $f_n$ of the form
	\begin{subequations}\label{pertubation scheme}
		\begin{align}
			g_n &= \sum_{k=1}^{\infty}\varepsilon^{2k-1}g_{n,2k-1} = \varepsilon g_{n,1}+ \varepsilon^3 g_{n,3} + \varepsilon^5 g_{n,5}+ \cdots ,\\
			h_n &= \sum_{k=1}^{\infty}\varepsilon^{2k-1}h_{n,2k-1} =\varepsilon h_{n,1}+ \varepsilon^3 h_{n,3} + \varepsilon^5 h_{n,5}+ \cdots  ,\\
			f_n &= 1+ \sum_{k=1}^{\infty}\varepsilon^{2k}f_{n,2k} =1+ \varepsilon^2 f_{n,2}+ \varepsilon^4 f_{n,4} + \varepsilon^6 f_{n,6}+ \cdots  ,
		\end{align}
	\end{subequations}
where $\varepsilon$ is a formal variable, used to track the number of building block functions in $g_{n,2k-1}$, $h_{n,2k-1}$ and $f_{n,2k}$. For soliton solution, the building blocks consist of exponential functions with different wave arguments. In fact, the form of $g_{n,1}$ and $h_{n,1}$ will be the sum of $N$ exponentials 
	\begin{equation*}
		E_j(n,t) =  e^{\eta_j(n,t)} = e^{np_j + tQ_j+ \theta_j}, \quad j = 1,\cdots, N,
	\end{equation*}
where the parameters $p_j, Q_j$ and $\theta_j$ are complex numbers denoting the wave number, angular frequency and phase constant, respectively. The computation proceeds by substituting the perturbation-like expansions \eqref{pertubation scheme} into the system of equations in \eqref{bilinear}, and collecting and solving for $g_{n,2k-1}$, $h_{n,2k-1}$ and $f_{n,2k}$ order by order in $\varepsilon$. If the expansions \eqref{pertubation scheme} truncate, one obtains the explicit formula for the soliton as a finite {sum} of products of exponentials $E_j(n,t)$.

In the remainder of this section we will show how { $g_{n}$, $h_{n}$ and $f_{n}$} are computed for a single fundamental soliton (see also \cite{Ohta00,[5]}), but also for a fundamental breather and a composite breather, via suitable parameter reductions informed by the spectral properties provided by the IST. In Section~2 we will iterate the algorithm to obtain ``2-body'' solutions, i.e., pairs of solitons or breathers.
If $g_n$ or $h_n$ is composed of $N$ solitons (with a fundamental or composite 1-breather counting as a sum of 2 solitons), then { we observe that the perturbation-like expansion \eqref{pertubation scheme}} terminates at order at most $2N$, allowing to build solutions with an arbitrary number of solitons and/or breathers of either type, following the parameter reductions illustrated below.

\subsection{One fundamental soliton (FS)} \label{ssec:FS1}
We show here how $g_n$, $h_n$ and $f_n$ are computed for the one FS solution. We provide details of the calculations, as the other solutions constructed in this section and the next can then be obtained analogously. { Substituting the expansions} \eqref{pertubation scheme} into the equations \eqref{bilinear} then { collecting} in order of $\varepsilon$, we get
\begin{subequations}\label{perturbation scheme}
        \begin{align}
			 &\mathcal{O}(\varepsilon): &&  i (\Delta h)^2 \frac{d g_{n,1}}{dt} + g_{n+1,1} - 2g_{n,1} + g_{n-1,1}  =  0, \,\label{order1a}\\
			 & && i (\Delta h)^2 \frac{d h_{n,1}}{dt} + h_{n+1,1} - 2h_{n,1} + h_{n-1,1} = 0, \label{order1b}\\
			 \nonumber \\ 
			 &\mathcal{O}(\varepsilon^2): && f_{n+1,2} -2f_{n,2} + f_{n-1,2}  = -\left(\Delta h\right)^2\left(g_{n,1}g^{*}_{n,1}+ h_{n,1}h^{*}_{n,1}\right)  ,\label{2nd order}\\[-10pt]
			 \nonumber \\
			 &\mathcal{O}(\varepsilon^3): &&  i (\Delta h)^2\frac{d g_{n,3}}{dt} - g_{n+1,3} + 2g_{n,3} - g_{n-1,3}  = i (\Delta h)^2 \left(\frac{d f_{n,2}}{dt} g_{n,1} - f_{n,2}\frac{d g_{n,1}}{dt}\right)  \qquad \qquad \quad\nonumber\\
			 & && \qquad\qquad\qquad\qquad\qquad\qquad\qquad\qquad\quad\quad
            + g_{n+1,1}f_{n-1,2} - 2g_{n,1}f_{n,2} +g_{n-1,1}f_{n+1,2} , \\
			 &  && i (\Delta h)^2\frac{d h_{n,3}}{dt} - h_{n+1,3} + 2h_{n,3} - h_{n-1,3}  = i (\Delta h)^2 \left( \frac{d f_{n,2}}{dt} h_{n,1} - f_{n,2}\frac{d h_{n,1}}{dt}    \right ) \qquad \qquad \,\, \nonumber\\
			 & && \qquad\qquad\qquad\qquad\qquad\qquad\qquad\qquad\quad\quad
             + h_{n+1,1}f_{n-1,2} - 2h_{n,1}f_{n,2} +h_{n-1,1}f_{n+1,2} , \\
			 \nonumber \\[-10pt]
			 &\mathcal{O}(\varepsilon^4): \quad && f_{n+1,4} - 2f_{n,4} + f_{n-1,4} = f_{n+1,2}f_{n-1,2} - f_{n,2}^2 -(\Delta h)^2\left(g_{n,1}g_{n,3}^{*} + g_{n,1}^{*}g_{n,3}\right)\,\nonumber\\ 
			 & && \qquad\qquad\qquad\qquad\qquad\qquad\qquad\qquad\qquad\qquad\!\!
             -(\Delta h)^2\left( h_{n,1}h_{n,3}^{*} + h_{n,1}^{*}h_{n,3}\right). 
		\end{align}
	\end{subequations}
	Since we seek 1-soliton solution, the forms of $g_{n,1}$ and $h_{n,1}$ consist of only 1 exponential function
	\begin{equation}\label{g1 1 soliton}
		g_{n,1} = \gamma_1 E_1(n,t) =  \gamma_1 e^{n p_1 + t Q_1 + \theta_1}, \qquad
		h_{n,1} = \gamma_2 E_1(n,t) = \gamma_2 e^{n p_1 + t Q_1 + \theta_1},
	\end{equation}
	where $\gamma_1$, $\gamma_2$, $p_1$, 
	$\theta_1$ and $Q_1$ are complex-valued parameters. 
	We observe that while the equations for $g_{n,1}$ and $h_{n,1}$ decouple, { all equations of higher order in $\varepsilon$ require knowledge of $g_{n,2k-1}$, $h_{n,2k-1}$, and $f_{n,2k}$ from previous orders}.
	Substituting $g_{n,1}$ and $h_{n,1}$ into their respective equation at order $\mathcal{O}(\varepsilon)$ yields:
	\begin{equation}\label{Q rule}
		Q_1 = Q(p_1) = \frac{-i \left(e^{p_1} -2 +e^{-p_1} \right)}{\Delta h^2} = -4i \frac{\sinh^2\left(\frac{p_1}{2}\right)}{\Delta h^2}.
	\end{equation}
	The right hand side of the equation at $\mathcal{O}(\varepsilon)$ suggests that $f_{n,2}$ should take the form 
	\begin{equation}\label{rule f2}
		f_{n,2} = c_1 E_1E_1^{*},
	\end{equation}
and upon substitution into \eqref{2nd order} gives: 
	\begin{equation}\label{coef f2}
		c_1 =  \frac{e^{p_1+ p_1^{*}}\big(\gamma_1\gamma_1^{*} + \gamma_2\gamma_2^{*}\big)\Delta h^2}{(e^{p_1+ p_1^{*}}-1)^2}.
	\end{equation}
We note that $p_1\in \C$ must be such that $\textrm{Re}(p_1)\ne 0$ for $c_1$ to be well defined.\footnote{In earlier works, e.g., \cite{[4]}, the analog of the $p_1$ parameter was restricted to $\mathrm{Re}(p_1)>0$, because the discrete eigenvalue to which it is associated was chosen to have modulus less than 1. This, however, is  not necessary, as choosing $\mathrm{Re}(p_1)<0$ would simply amount to switching the eigenvalues in an octet.} 
	Proceeding to the next order, we see that $g_{n,3}$ must consist of { the sum of exponential functions resulting from the products of $g_{n,1}f_{n,2}$ as well as the combinations of $g_{n\pm 1,1}f_{n\pm1,2}$. Moreover, due to the definition of $E_j(n,t)$, a shift in $n$ yields a scalar multiple of $E_j(n,t)$. Thus, we obtain:}
	\begin{align}\label{rule g3}
		g_{n,3} &= c_2 E_1^2E_1^{*},
	\end{align}
which in turn implies:
    \begin{align*}
 		i (\Delta h)^2 &\left(\frac{d f_{n,2}}{dt} g_{n,1} - f_{n,2}\frac{d 
 		g_{n,1}}{dt}\right) + g_{n+1,1}f_{n-1,2} - 2g_{n,1}f_{n,2} +g_{n-1,1}f_{n+1,2} \\
 	&\qquad=\left(i \Delta h^2 Q_1^{*} + e^{p_1^{*}} - 2 + e^{-p_1^{*}}\right)c_1\gamma_1E_1^2E_1^{*} \\
 	&\qquad=\left[i \Delta h^2 \frac{i \left(e^{p_1^{*}} -2 +e^{-p_1^{*}} \right)}{\Delta h^2} + \left( e^{p_1^{*}} - 2 + e^{-p_1^{*}}\right)\right]c_1\gamma_1E_1^2E_1^{*}  =0,
 	\end{align*}
 which implies that the coefficient $c_2$ of $g_{n,3}$ must be zero. Then, it follows by symmetry that $h_{n,3}$ is zero. Similarly, at order $\mathcal{O}(\varepsilon^4)$, we can easily check that the right hand side is zero which forces $f_{n,4}$ to be zero. Consequently, $g_{n,5}$, $h_{n,5}$ are zeros, and so is $f_{n,6}$, etc. Thus, the expansions of $g_n$, $h_n$ and $f_n$ truncate. Setting $\varepsilon=1$ in { the expansions} \eqref{pertubation scheme} we obtain:
	\begin{align}
		g_{n,1} = \gamma_1 E_1(n,t), \qquad 
		h_{n,1} = \gamma_2E_1(n,t), \qquad
		f_{n,2} = 1+ \frac{e^{p_1+ p_1^{*}}\big(\gamma_1\gamma_1^{*} + \gamma_2\gamma_2^{*}\big)\Delta h^2}{(e^{p_1+ p_1^{*}}-1)^2} E_1(n,t)E_1^{*}(n,t).
	\end{align} 
Note that $f_{n,2}>0$ for all $n\in \Z$ and $t\in \R$, and therefore the one FS solution is regular for any choice of the parameters, and its explicit form is given by:
\begin{equation}\label{FS1 explicit}
 		\bd q_n(t) = \frac{e^{np_1+ t Q_1+ \theta_1}}{1+ \frac{e^{p_1+ p_1^{*}}\big(\gamma_1\gamma_1^{*} + \gamma_2\gamma_2^{*}\big)\Delta h^2}{(e^{p_1+ p_1^{*}}-1)^2} e^{2\textrm{Re}(np_1+ tQ_1+ \theta_1)}}\left(\begin{array}{c}
 			\gamma_1 \\ \gamma_2
 		\end{array}\right),
 \end{equation}
 where 
 \begin{gather*}
 	Q_1 =  -\frac{4i}{\Delta h^2} \sinh^2\left(\frac{p_1}{2}\right),\quad 
 	p_1 \in \C, \quad \textrm{Re}(p_1)\ne 0,\quad
 	\gamma_1, \gamma_2, \theta_1 \in \C.
 \end{gather*} 
Equation~\eqref{FS1 explicit} reduces to the one given in \cite{TUW99,AOT99,[4]}. Indeed, for $\theta_1 = 2a_1 + \ln\left(\frac{2\sinh^2(2a_1)}{\| \bd \gamma\|^2}\right) + i \left(\pi + 2b_1 \right) $, we can recast it as:
\begin{equation}
	\bd q_n(t) =  - \sinh(2a_1)\sech(\zeta - d_1)e^{i\left[2b_1(n+1)- 2\tau (\omega-1)\right]}\frac{\bd \gamma}{\| \bd \gamma\|}
\end{equation}
where 
 \begin{gather}
 	\tau = \frac{t}{\Delta h^2}, \qquad  \bd \gamma =\begin{pmatrix}\gamma_1 \\ \gamma_2\end{pmatrix}, \qquad d_1 = \ln \frac{\|\bd \gamma \|}{\sinh(2a_1)}, \qquad \zeta=  2a_1[(n+1)- v_1 \tau],\\
 	\omega =  \cosh(2a_1)\cos(2b_1), \qquad  v_1 = -\frac{1}{a_1}\sinh(2a_1)\sin(2b_1).
 \end{gather}
    
\subsection{One fundamental breather (FB)} \label{ssec:FB1}
{In the} subsequent sections, for the sake of simplicity, we will omit the free parameter $\theta_j$, denoting a phase shift, in the exponential function $E_j(n,t)= np_j + tQ_j + \theta_j$, unless our computation calls for $\theta_j$ to be defined explicitly.

\begin{figure}[H]
		\centering
		\includegraphics[width=0.7 \linewidth]{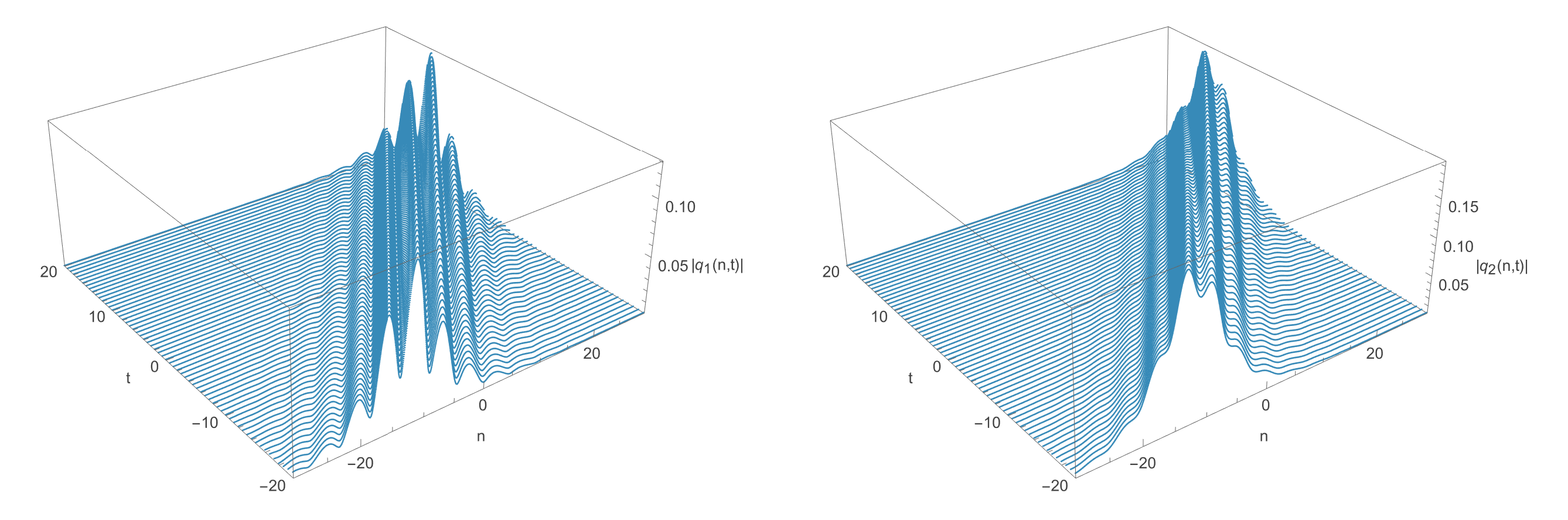}
		\caption{Fundamental breather with $p_1 = 1/5 - i2\pi/3$ and $\bd C = \left(\begin{array}{cc}
				2 & 8\\
				1 & 4
			\end{array}\right)$. Left: $|q^{(1)}_n(t)|$. Right:  $|q^{(2)}_n(t)|$.}
		\label{fig:1FB}
	\end{figure}

In \cite{[4],APT04b,APT06,13} a single FB solution {is} obtained by choosing the norming constant $\bd C$ of rank 1, with two columns proportional to each other:
		\begin{equation}\label{FB norming constant}
			\bd C = \left(\begin{array}{cc}
				\mu \gamma_1 & \kappa \gamma_1\\
				\mu \gamma_2 & \kappa \gamma_2
			\end{array} \right), \quad \bd \gamma =\begin{pmatrix}\gamma_1 \\ \gamma_2\end{pmatrix} \in \C^2,\quad \mu, \kappa\in \C.
		\end{equation}
It { is} also observed that a FB is a superposition of two orthogonally polarized fundamental solitons which share the same amplitude and velocity, but have opposite carrier frequencies. Let us then define  $\eta_j(n,t):= np_j + t Q_j$ and denote its real part by $\xi_j(n,t)$ and its imaginary part by $\nu_j(n,t)$, so the FS can be written as: 
	\begin{gather*}
		\bd q_n(t) 
		=  \frac{c_1^{-1}e^{\xi_1(n,t)}e^{i\nu_1(n,t)}}{c_1^{-1}+ e^{2\xi_1(n,t)}}\bd \gamma,
	\end{gather*}
where $c_1$ is defined in Eq.~\eqref{coef f2}. 
Let $$\widetilde{E}_1(n,t)= n \widetilde{p}_1+ t \widetilde{Q}_1 \equiv \widetilde{\xi}_1+ i \widetilde{\nu}_1,$$
so that $\widetilde{\xi}_1=\mathrm{Re}({\widetilde{E}_1})$, $\widetilde{\nu}_1=\mathrm{Im} ({\widetilde{E}_1})$, be the exponential function associated with the orthogonally polarized soliton. Then, the soliton and its orthogonally polarized companion have the same amplitude and velocity if we impose
		\begin{align}\label{tilde condition}
			\xi_1(n,t) =  \widetilde{\xi}_1(n,t),
		\end{align}
	which is equivalent to $\textrm{Re}(p_1) =  \textrm{Re}(\widetilde{p}_1)$, and $\textrm{Re}(Q_1) = \textrm{Re}(\widetilde{Q}_1)$. 
	We repeat the same computation as in Section \ref{ssec:FS1} with the ansatz for $g_{n,1}$ and $h_{n,1}$ modified as follows
    \begin{align}
		g_{n,1} &= A_1 E_1(n,t) + A_2  \widetilde{E}_1(n,t), \label{FB1 ansatz1}\\
		h_{n,1} &= B_1 E_1(n,t) + B_2   \widetilde{E}_1(n,t). \label{FB1 ansatz2}
	\end{align}
    Similarly to the fundamental soliton case, the coefficients $(A_i, B_i)^T$ for $i= 1,2$ are related to the columns of the norming constant $\bd C$ in \eqref{FB norming constant}. We observe that the forms of $g_{n,1}$ and $h_{n,1}$ in equations \eqref{FB1 ansatz1} and \eqref{FB1 ansatz2} imply that the fundamental solution set for the system at the first order $\varepsilon$ is given by 
    \begin{align*}
    \Phi = \left\{ E_1\left( \begin{array}{c} A_1  \\ B_1\end{array}\right), \widetilde{E}_1\left( \begin{array}{c} A_2 \\ B_2 \end{array}\right)\right \}.
    \end{align*}
    However, $\bd C$ is of rank 1, containing only one linearly independent column. Hence, to have a linearly independent fundamental solution set $\Phi$, we take $(A_2, B_2)^{T}$ to be the orthogonal of the second column in $\bd C$. That is, $(A_1, B_1)^T = \mu^{*}(\gamma_1^{*}, \gamma_2^{*})^T$ and $(A_2,B_2)^T =  \kappa(-\gamma_2, \gamma_1)^{T}$. \\
At order $\mathcal{O}(\varepsilon)$ of the perturbation-like scheme, Eqs.~\eqref{order1a} and \eqref{order1b}, we see that for $p_1 =  2a_1 + i 2b_1$, the condition \eqref{tilde condition} is satisfied if
	\begin{equation}
		\widetilde{p}_1 =  2a_1 + i (\pi - 2b_1), \qquad
		\widetilde{Q}_1 =   -\frac{4i}{\Delta h^2} \sinh^2\left(\frac{\widetilde{p}_1}{2}\right).
	\end{equation}
Then the second order of $\varepsilon$ in the expansion yields
	\begin{align}
		f_{n,2} &= \frac{e^{p_1+ p_1^{*}}|\mu|^2\| \bd \gamma \|^2\Delta h^2}{(e^{p_1+ p_1^{*}}-1)^2} E_1 E_1^{*} + \frac{e^{\widetilde{p}_1+ \widetilde{p}_1^{*}} |\kappa|^2 \|\bd \gamma^{\perp} \|^2\Delta h^2}{(e^{p_1+ p_1^{*}}-1)^2} \widetilde{E}_1\widetilde{E}_1^{*},
	\end{align}
	and all subsequent orders { of} $\varepsilon$ vanish: $g_{n,2k-1}=h_{n,2k-1}=0$ and $f_{n,2k}=0$ for all $k\in \mathbb{N}$ with $k\ge 2$.
	Hence, the expression of a 1 FB solution is given by
	\begin{align}
		\bd q_n(t) &= \frac{1}{1+ f_{n,2} }\left(\begin{array}{c}
			g_{n,1}\\ h_{n,1}
		\end{array}\right) \notag \\
	&=  \frac{  \left(\begin{array}{cc} \mu^{*} \gamma_1^{*} & -\kappa \gamma_2 \\ \mu^{*}\gamma_2^{*} & \kappa \gamma_1 \end{array}\right) \left( \begin{array}{c} E_1\\ \widetilde{E}_1 \end{array}\right) }{1+ \frac{e^{p_1+ p_1^{*}}|\mu|^2\| \bd \gamma \|^2\Delta h^2}{(e^{p_1+ p_1^{*}}-1)^2} E_1E_1^{*} + \frac{e^{\widetilde{p}_1+ \widetilde{p}_1^{*}}|\kappa|^2 \|\bd \gamma^{\perp} \|^2\Delta h^2}{(e^{p_1+ p_1^{*}}-1)^2} \widetilde{E}_1\widetilde{E}_1^{*} },
	\end{align}
and it is plotted in Fig.~\ref{fig:1FB} for some specific choices of parameters.

\subsection{Composite breather (CB)}\label{ssec:CB1}
    Compared to the 1 FB solution, the CB is also a superposition of two fundamental solitons, however, the corresponding norming constant is full rank, i.e.: 
	\begin{equation}\label{CB norming constant}
		\bd C = \left(\begin{array}{cc}
			\gamma_1 & \epsilon_1\\
			\gamma_2 & \epsilon_2
		\end{array}
		\right) = \left( \bd \gamma, \bd \epsilon \right)
	\end{equation} 
where $\bd \gamma = (\gamma_1, \gamma_2)^T$ and $\bd \epsilon= (\epsilon_1, \epsilon_2)^T$ are linearly independent vectors in $\C^2$.	In this case, $g_{n,1}$ and $h_{n,1}$ take the form 
\begin{gather}\label{CB- ansatz}
		g_{n,1} = \gamma_1^{*} E_1 + \epsilon_1  \widetilde{E}_1,\qquad 
		h_{n,1} = \gamma_2^{*} E_1 + \epsilon_2   \widetilde{E}_1,
\end{gather}
where $E_1$ and $\widetilde{E}_1$ are the same as in the FB solution.	Solving for $g_{n,2k-1}$, $h_{n,2k-1}$ and $f_{n,2k}$ order-by-order in $\varepsilon$ using the equations in the perturbation-like expansions, we get
\begin{align}\label{CB-soln}
		\bd q_n(t) &= \frac{1}{1+ f_{n,2}+ f_{n,4} }\left(\begin{array}{c}
			g_{n,1} +g_{n,3}\\ h_{n,1}+h_{n,3}
		\end{array}\right)
	\end{align}
	where $g_{n,1}$ and $h_{n,1}$ are given in Eq.~\eqref{CB- ansatz}, and
	\begin{align}
		g_{n,3} &= \beta_1 E_1E_1^{*}\widetilde{E}_1 + \beta_2 E_1\widetilde{E}_1\widetilde{E}_1^{*},\\
		h_{n,3} &=  \delta_1 E_1E_1^{*}\widetilde{E}_1 + \delta_2 E_1\widetilde{E}_1\widetilde{E}_1^{*},\\
		f_{n,2} &=\alpha_1 E_1 E_1^{*}  + \alpha_2 E_1\widetilde{E}_1^{*} + \alpha_3 \widetilde{E}_1E_1^{*}+ \alpha_4 \widetilde{E}_1\widetilde{E}_1^{*},\\
		f_{n,4} &=\rho_1  E_1E_1^{*}\widetilde{E}_1\widetilde{E}_1^{*},
	\end{align}
with
\begin{subequations}
    \begin{gather}
		\alpha_1 = A_{(1, 1^{*})} \|\bd \gamma \|^2,  \quad \alpha_2 = A_{(1, \widetilde{1}^{*})} \langle \bd \gamma^{*}, \bd \epsilon^{*} \rangle, \quad \alpha_3 = A_{(\widetilde{1}, 1^{*})} \langle \bd \gamma, \bd \epsilon \rangle, \quad \alpha_4 = A_{(\widetilde{1}, \widetilde{1}^{*})} \| \bd \epsilon \|^2 \\
		\beta_1 = P_{(1, \widetilde{1})}\left( \gamma_1^{*} \alpha_3 P_{(1, 1^{*})} - \epsilon_1 \alpha_1 P_{(\widetilde{1}, 1^{*})}    \right) , \quad \beta_2 = P_{(1, \widetilde{1})}\left( \gamma_1^{*} \alpha_4 P_{(1, \widetilde{1}^{*})} - \epsilon_1 \alpha_2 P_{(\widetilde{1}, \widetilde{1}^{*})}    \right),  \\
		\delta_1 = P_{(1, \widetilde{1})}\left( \gamma_2^{*} \alpha_3 P_{(1, 1^{*})} - \epsilon_2 \alpha_1 P_{(\widetilde{1}, 1^{*})}    \right) , \quad  \delta_2 = P_{(1, \widetilde{1})}\left( \gamma_2^{*} \alpha_4 P_{(1, \widetilde{1}^{*})} - \epsilon_2 \alpha_2 P_{(\widetilde{1}, \widetilde{1}^{*})}    \right), \\
		\rho_1 =  A_{(1,1^*,\widetilde{1},\widetilde{1}^{*})}\left[\left(\gamma_1^* \beta_2^*+ \gamma_2^*\delta_2^* + \epsilon_1 \beta_1^* + \epsilon_2 \delta_1^* \right) + \left( \gamma_1 \beta_2 + \gamma_2\delta_2 + \epsilon_1^{*} \beta_1+ \epsilon_2^{*}\delta_1 \right) \right] \qquad\,\,\nonumber \\
		 \qquad\qquad\qquad\qquad\qquad\qquad\qquad\qquad\quad\quad - S^2_{[(1,\widetilde{1}^*), (\widetilde{1},1^*)]}\alpha_2\alpha_3 - S^2_{[(1,1^*),(\widetilde{1},\widetilde{1}^*)]}\alpha_1\alpha_4, \\
		A_{(j,k^*)} = \frac{e^{p_j+p_k^{*}} \Delta h^2}{\left(e^{p_j+p_k^{*}}-1\right)^2}, \quad 
        P_{(j,k^*)} = \frac{e^{p_j} + e^{p_k^{*}}}{e^{p_j+p_k^{*}}-1}, \quad 
        P_{(1,\widetilde{1})} = \frac{e^{p_1} - e^{\widetilde{p}_1}}{e^{p_1+ \widetilde{p}_1}+1},
        \quad {j, k}\in \{1,\widetilde{1} \}, \label{coeff func1} \\
		S_{[(j,k^*),(l,m^*)]} = \frac{e^{p_j+ p_k^*}- e^{p_l+ p_m^*}}{e^{p_j+p_k^*+p_l+p_m^*}-1}, \quad {j,k,l,m} \in \{1,\widetilde{1}\}. \label{coeff func2}
	\end{gather}
    \end{subequations}
In the above coefficients, $\langle \mathbf{u}, \mathbf{v}\rangle=\sum_{j=1,2}u_jv_j$ denotes the inner product of 2-component vectors $\mathbf{u}, \mathbf{v}$.
The subscripts of the auxiliary functions in Eqs.~\eqref{coeff func1} and \eqref{coeff func2} denote the contributions of the exponential functions from which a particular coefficient is to be computed. For example, { the coefficients $\alpha_j$} computed at order $\varepsilon^2$ in the perturbation-like scheme { contain the term $A_{(j,k*)}$ which consists} of the contributions from the wave numbers $p_j$ and $p_k^{*}$ of the exponential functions $E_j$ and $E_k^{*}$, while at order $\varepsilon^4$, since $f_{n,4}$ is a product of all building block exponential functions ($E_1$, $\widetilde{E}_1$ and their conjugates), the coefficient $\rho_1$ has a factor of $A_{(1,1^*,\widetilde{1},\widetilde{1}^{*})}$ which indicates the ``mixing'' of all building block exponential functions. 
    
 \smallskip   

\begin{figure}[H]
		\centering
		\includegraphics[width=0.7 \linewidth]{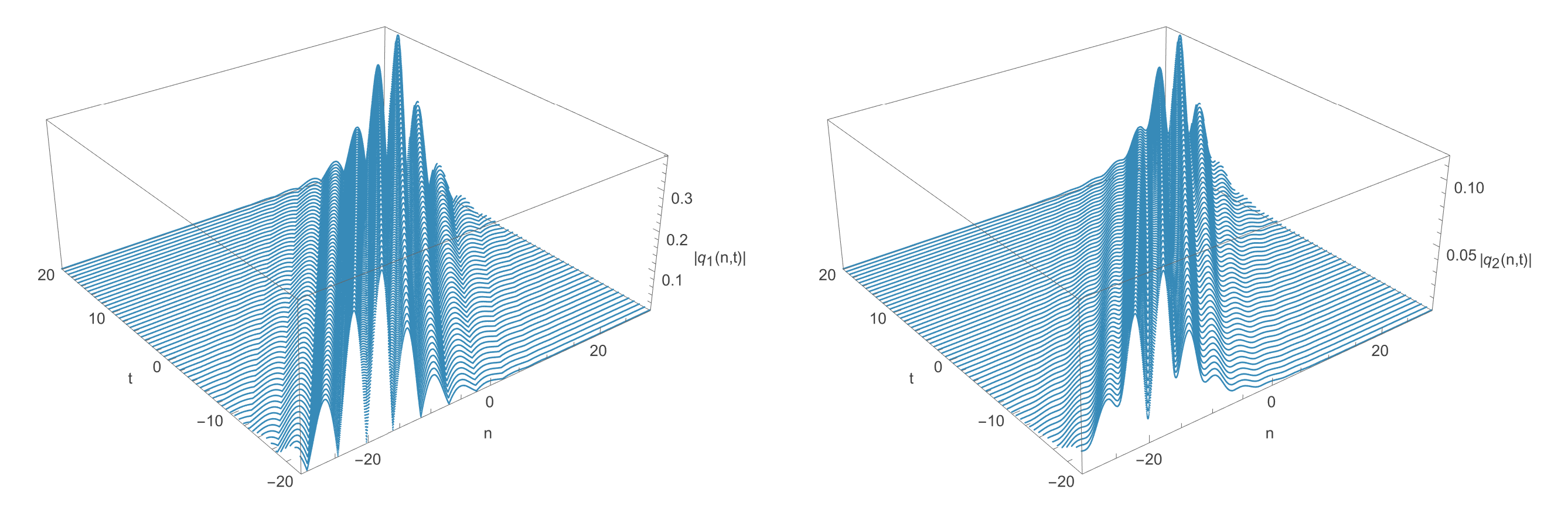}
		\caption{Composite breather with $p_1 = \frac15 - i \frac{2\pi}{3}$ and $\bd C = \left(\begin{array}{cc}
				2 & 2\\
				1/5 & 1
			\end{array}\right)$. Left: $|q^{(1)}_n(t)|$. Right: $|q^{(2)}_n(t)|$.}
		\label{fig:1soliton-cb}
	\end{figure}

\begin{figure}[H]
		\centering
		\includegraphics[width=0.7 \linewidth]{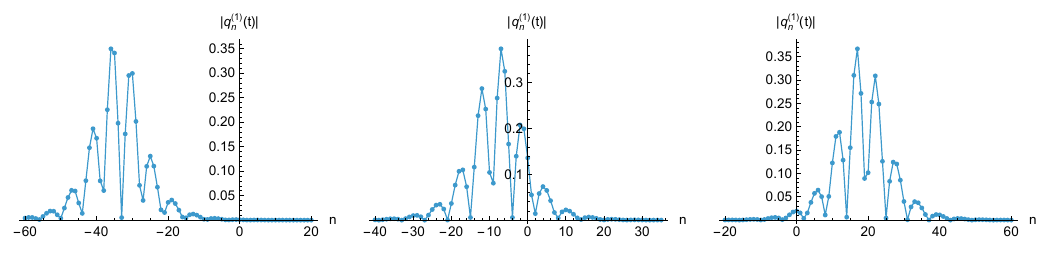}
		\includegraphics[width=0.7 \linewidth]{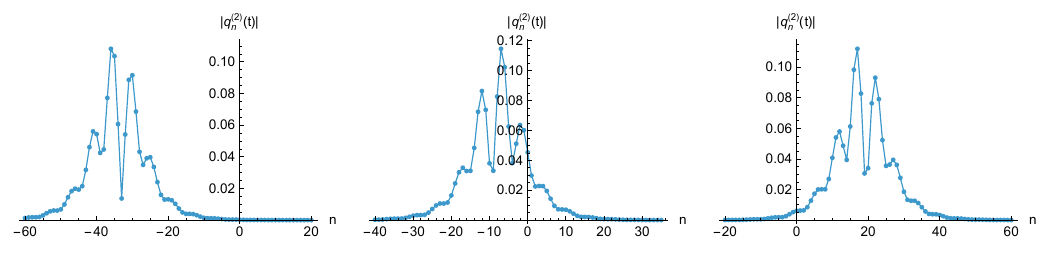}
		\caption{Snapshots of $|q^{(1)}_n(t) |$ (top) and $|q^{(2)}_n(t) |$ (bottom) of the CB in Fig~\ref{fig:1soliton-cb} at $t= -15, 0$, and $15$.}
		\label{fig:cb-q1-snapshot}
	\end{figure}

Plots of a CB solution are provided in Figs.~\ref{fig:1soliton-cb} and \ref{fig:cb-q1-snapshot}, for specific choices of the parameters.

\section{Exact ``2-body'' solutions and their long-time asymptotics}
In this section, we will use the Hirota's formalism to obtain explicit expressions of solutions exhibiting 2 coherent structures (solitons, breathers or soliton-breather), and compute their long-time asymptotics to describe the interactions. We recall the formulae for the building block exponential functions $E_j$, $\widetilde{E}_j$, and their parameters for $j\in \mathbb{N}$, and $a_j, b_j\in \R$ with $a_j>0$:
	\begin{subequations}
		\begin{gather}
			E_j (n,t) = e^{\eta_j(n,t)} = e^{np_j+ tQ_j}= e^{\xi_j (n,t)+ i\nu_j(n,t)}, \label{def E} \\
			\widetilde{E}_j (n,t) = e^{\widetilde{\eta}_j(n,t)} =  e^{n\widetilde{p}_j+ t\widetilde{Q}_j}= e^{\xi_j(n,t)+ i\widetilde{\nu}_j(n,t)},\\
		Q_j =  \frac{-ie^{-p_j}\left(e^{p_j}-1\right)^2}{\Delta h^2}, \label{def Q} \quad 
		\widetilde{Q}_j =  \frac{-ie^{-\widetilde{p}_j}\left( e^{\widetilde{p}_j}-1\right)^2}{\Delta h^2},\\
         p_j = 2a_j + i 2b_j, \label{def p} \quad\quad \quad\quad   
		\widetilde{p}_j = 2a_j + i (\pi - 2b_j).
		\end{gather}
	\end{subequations}
Note that $\eta_j(n,t)$ and $\widetilde{\eta}_j(n,t)$ have the same real part, and $^*$ denotes complex conjugation { as before}.
	\subsection{Fundamental soliton - fundamental soliton (FS-FS)}
Let $E_1(n,t)$ be the exponential function associated with the first soliton and $E_2(n,t)$ the second soliton.
The corresponding norming constant matrices are given as
		\begin{align}
			\bd C_1 =  \left(\begin{array}{cc}
				\gamma_1 & 0\\ \gamma_2 & 0
			\end{array}\right), \qquad \bd C_2 =  \left(\begin{array}{cc}
			\kappa_1 & 0\\ \kappa_2 & 0
		\end{array}\right),\quad  \gamma_j, \kappa_j \in \C, \quad j \in \{1,2\}.
		\end{align}
Then, the exact FS-FS solution has the form:
	\begin{equation}
	 	\bd q_n(t) = \left(\begin{array}{c}
			q^{(1)}_n(t)\\ q^{(2)}_n(t)\end{array}\right) = \frac{1}{1+ f_{n,2}+ f_{n,4}}\left(\begin{array}{c}
	 			g_{n,1}+ g_{n,3}\\ h_{n,1}+h_{n,3}
	 		\end{array}\right)
	\end{equation}
where
	\begin{subequations}\label{defFS-FS}
		\begin{gather}
			g_{n,1} = \gamma_1^{*}E_1 + \kappa_1^{*}E_2, \quad 
			h_{n,1} = \gamma_2^{*}E_1 + \kappa_2^{*}E_2,\\
			g_{n,3} = \beta_1 E_1E_1^{*}E_2 +  \beta_2 E_1E_2E_2^*,\quad
			h_{n,3} = \delta_1 E_1E_1^{*}E_2 +  \delta_2 E_1E_2E_2^*,\\
            f_{n,2} = \alpha_1 E_1 E_1^{*}  + \alpha_2 E_1E_2^{*} + \alpha_3 E_2E_1^{*}+ \alpha_4 E_2E_2^{*}, \quad 
			f_{n,4} = \rho_1  E_1E_1^{*}E_2E_2^{*}, 
		\end{gather}
\end{subequations}
and
\vspace{-.3cm}
\begin{subequations}
\label{coeffs_FSFS}
    \begin{gather}
		\bd \gamma = \left(\begin{array}{c}
			\gamma_1 \\ \gamma_2 
		\end{array}\right),\quad    \bd \kappa = \left(\begin{array}{c}
		\kappa_1 \\ \kappa_2 \end{array}\right),\\
		\alpha_1 = \langle\bd \gamma, \bd\gamma^* \rangle A_{(1,1^*)},\quad \alpha_2 =  \langle\bd \gamma, \bd\kappa^* \rangle A_{(1, 2^*)},\quad \alpha_3 =  \langle\bd \kappa, \bd\gamma^* \rangle A_{(2,1^*)},\quad \alpha_4 =  \langle\bd \kappa, \bd\kappa^* \rangle A_{(2,2^*)},\\
		\beta_1 = P_{(1,2)}P_{(1, 1^*)}\alpha_3\gamma_1^{*} -
		P_{(1,2)}P_{(2,1^*)}\alpha_1\kappa_1^*,\quad 
		\beta_2 = P_{(1,2)}P_{(1,2^*)}\alpha_4\gamma_1^* - P_{(1,2)}P_{(2,2^*)}\alpha_2\kappa_1^*,\\
		\delta_1	= P_{(1,2)}P_{(1, 1^*)}\alpha_3\gamma_2^{*} -
			P_{(1,2)}P_{(2,1^*)}\alpha_1\kappa_2^*,\quad 
		\delta_2	= P_{(1,2)}P_{(1,2^*)}\alpha_4\gamma_2^* - P_{(1,2)}P_{(2,2^*)}\alpha_2\kappa_2^*,\\
			\rho_1 = A_{(1,1^*,2,2^*)}\left(\frac{}{}\gamma_1^* \beta_2^*+ \gamma_2^*\delta_2^* + \kappa_1^* \beta_1^*+ \kappa_2^*\delta_1^*+ \gamma_1 \beta_2+ \gamma_2\delta_2 + \kappa_1 \beta_1+ \kappa_2\delta_1 \right) \nonumber\qquad\qquad\qquad\\ 
			  \qquad\qquad\qquad\qquad\qquad\qquad\qquad\qquad\qquad\qquad
              - S^2_{[(1,2^*),(2,1^*)]}\alpha_2\alpha_3- S^2_{[(1,1^*),(2,2^*)]}\alpha_1\alpha_2,
              \\ 
			A_{(j,k^*)} = \frac{e^{p_j+p_k^{*}} \Delta h^2}{\left(e^{p_j+p_k^{*}}-1\right)^2}, \quad 
            P_{(j,k^*)} = \frac{e^{p_j} + e^{p_k^{*}}}{e^{p_j+p_k^{*}}-1},\quad P_{(1,2)} = \frac{e^{p_1} - e^{p_2}}{e^{p_1+p_2}+1},
            \quad {j, k}\in \{1,2\},\\
			S_{[(j,k^*),(l,m^*)]}= \frac{e^{p_j+ p_k^*}- e^{p_l+ p_m^*}}{e^{p_j+p_k^*+p_l+p_m^*}-1}, \quad j,k,l,m \in \{1,2\}.
	\end{gather}
\end{subequations}    
\paragraph{Long-time asymptotics of a FS-FS solution.}
In order to study the long-time asymptotics of the interaction, we substitute {the} definitions of $E_j$ in Eq.~\eqref{def E} into Eqs.~\eqref{defFS-FS}
	\begin{subequations}
		\begin{gather}
			g_{n,1} = \gamma_1^{*}e^{\xi_1 + i \nu_1} + \kappa_1^{*}e^{\xi_2 + i \nu_2},\quad 
			h_{n,1} = \gamma_2^{*}e^{\xi_1 + i \nu_1} + \kappa_2^{*}e^{\xi_2 + i \nu_2},\qquad\\
			g_{n,3} = \beta_1 e^{2\xi_1+ \xi_2 + i \nu_2}  +  \beta_2 e^{2\xi_2+ \xi_1 + i \nu_1},\quad
			h_{n,3} = \delta_1 e^{2\xi_1+ \xi_2 + i \nu_2} +  \delta_2 e^{2\xi_2+ \xi_1 + i \nu_1},\\
            f_{n,2} = \alpha_1 e^{2\xi_1}  + \alpha_2 e^{\xi_1 +\xi_2 + i (\nu_1- \nu_2)} + \alpha_3e^{\xi_1 +\xi_2 + i (\nu_2- \nu_1)}+ \alpha_4 e^{2\xi_2},\quad
			f_{n,4} = \rho_1  e^{2\xi_2+ 2\xi_1}.
		\end{gather}
	\end{subequations}
Moreover, from the definitions of $p_j$, $Q_j$ in Eqs.~\eqref{def p} and \eqref{def Q} we have
\begin{equation}
	 \xi_j:=\textrm{Re}(\eta_j) \equiv  n \textrm{ Re}(p_j)+ t\textrm{ Re}(Q_j)
	= 2na_j + 2t \sinh(2a_j)\sin(b_j).
\end{equation}
Let $v_j = -\frac{1}{a_j}\sinh(2a_j)\sin(2b_j)$ denote the soliton velocities, and recast $\xi_j$ as
	\begin{align}
    \label{e:xij}
		\xi_j =  2a_j \left(n- v_j t\right), \qquad j=1,2.
	\end{align}
We then assume without loss of generality $v_1 < v_2$, consider the reference frame of the FS associated with $\xi_1$, and write:
	\begin{equation}
    \label{e:xi2vsxi1}
		\xi_2 = \frac{a_2}{a_1} \xi_1 - 2a_2 \left(v_2 -v_1\right) t.
	\end{equation}
Since $v_2- v_1 >0$, $\xi_2  > 0$ as $t \to -\infty$. That is, $e^{\xi_2}\to \infty $, or, equivalently, $e^{-\xi_2}\to 0$ as $t \to -\infty$. Hence, the leading terms of $\bd q (n,t)$ as $t\to -\infty$ are given by
	\begin{subequations}
		\begin{align}
			q^{(1)}_n(t) &\sim \frac{\kappa_1^{*}e^{\xi_2 + i \nu_2}+ \beta_2 e^{2\xi_2+ \xi_1+ i\nu_1}}{1 + \alpha_4 e^{2\xi_2} + \rho_1 e^{2\xi_1+2\xi_2} }
			= \frac{e^{2\xi_2}\left(\kappa_1^{*}e^{-\xi_2 + i \nu_2}+ \beta_2 e^{\xi_1+ i\nu_1} \right)}{e^{2\xi_2}\left(e^{-2\xi_2}+ \alpha_4 +\rho_1 e^{2\xi_1} \right)}
			\sim \frac{ \beta_2 e^{\xi_1+ i\nu_1}}{\alpha_4 +\rho_1 e^{2\xi_1}},\\
			q^{(2)}_n(t) &\sim \frac{ \delta_2 e^{\xi_1+ i\nu_1}}{\alpha_4 +\rho_1 e^{2\xi_1}},
		\end{align}
	\end{subequations}
where the coefficients are given explicitly in Eqs.~\eqref{coeffs_FSFS}.
In the limit $t \to \infty$, then $\xi_2 <0$. That is, $e^{\xi_2}\to 0 $ as $t \to \infty$. Then, by similar computation, the leading terms of $\bd q (n,t)$ are
	\begin{gather}
		q^{(1)}_n(t) \sim \frac{\gamma_1^{*} e^{\xi_1+i \nu_1} }{1+ \alpha_1 e^{2\xi_1} },\quad 
		q^{(2)}_n(t) \sim \frac{\gamma_2^{*} e^{\xi_1+i \nu_1} }{1+ \alpha_1 e^{2\xi_1} }.
	\end{gather}
On the other hand, if the long-time asymptotics is computed in the co-moving frame of the second FS, i.e., keeping $\xi_2=\textrm{const}$, then
\begin{equation}
\label{e:xi1vsxi2}
		\xi_1 = \frac{a_1}{a_2}\xi_2 - 2a_1(v_1-v_2)t.
\end{equation}
For $t\to -\infty$, since $v_2>v_1$, $\xi_1<0$. That is, $e^{\xi_1}\to 0$ as $t \to -\infty$. The leading terms of $\bd q(n,t)$ are 
\begin{subequations}
\begin{gather}
		q^{(1)}_n(t) \sim \frac{\kappa_1^{*}e^{\xi_2+ i\nu_2}}{1+ \alpha_4 e^{2\xi_2}}, \quad
		q^{(2)}_n(t) \sim \frac{\kappa_2^{*}e^{\xi_2+ i\nu_2}}{1+ \alpha_4 e^{2\xi_2}}.
\end{gather}
For $t\to \infty$, since $v_2>v_1$, $\xi_1 > 0$. That is, $e^{\xi_1}\to \infty$ as $t \to \infty$. The leading terms of $\bd q(n,t)$ are 
\begin{gather}
		q^{(1)}_n(t) \sim \frac{\gamma_1^{*}e^{\xi_1 + i\nu_1}+ \beta_1 e^{2\xi_1+ \xi_2 + i\nu_2}}{1+ \alpha_1 e^{2\xi_1}+ \rho_1 e^{2\xi_1+ 2\xi_2}} 
		\sim \frac{\beta_1 e^{\xi_2 + i\nu_2}}{ \alpha_1+ \rho_1 e^{2\xi_2}},\quad 
		q^{(2)}_n(t) \sim \frac{\delta_1 e^{\xi_2 + i\nu_2}}{ \alpha_1+ \rho_1 e^{2\xi_2}},
\end{gather}
\end{subequations}
where again the coefficients are given explicitly in Eqs.~\eqref{coeffs_FSFS}. 

\begin{figure}[H]
		\centering
		\includegraphics[width=0.7\linewidth]{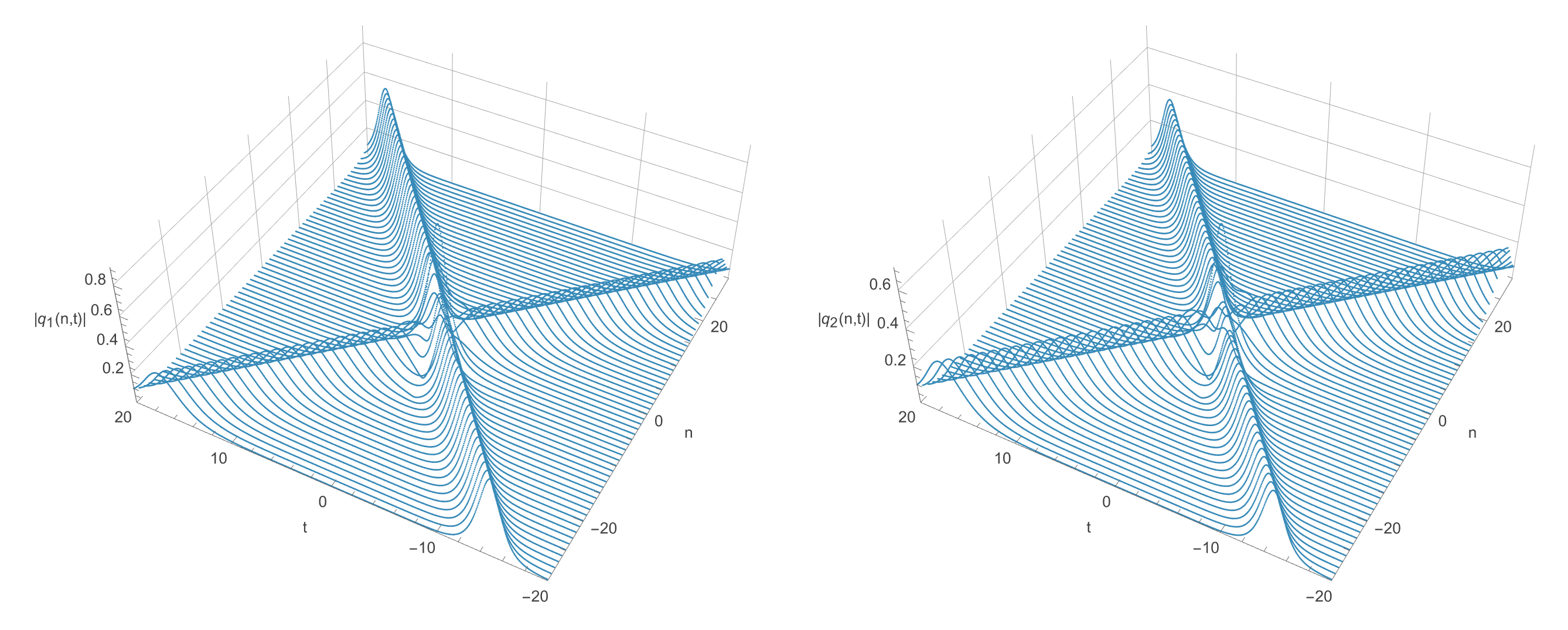}
		\includegraphics[width=0.7\linewidth]{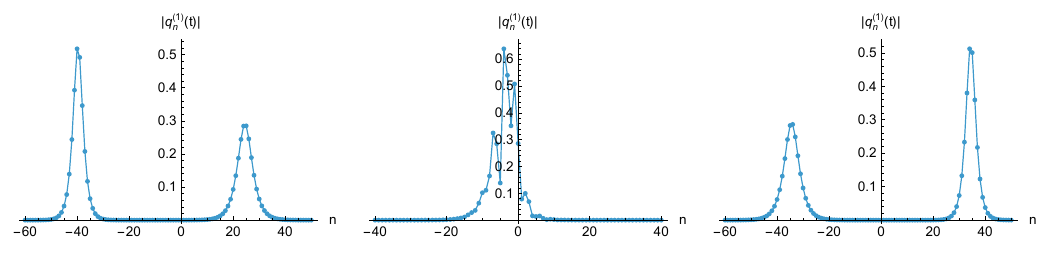}
		\includegraphics[width=0.7\linewidth]{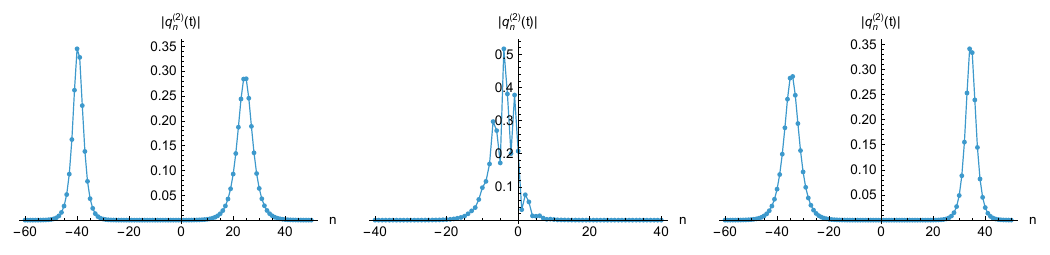}
		\caption{Top panel: Fundamental soliton-fundamental soliton solution with $p_1 = 2/5+ i \pi/4 $, $\bd \gamma = (5,5)^T$, $p_2 = 3/5 -i\pi/3 $, $\bd \kappa = (6,4)^T$. Middle and bottom panel: Snapshots of $|q^{(1)}_n(t)|$ and $|q^{(2)}_n(t)|$ respectively at $t =  -20, 0 , 20.$}
        \label{fig:fs-fs-interaction}
	\end{figure}

The above formulas confirm that upon interaction each FS retains its nature and velocity, while generically exhibiting shifts in both the soliton center and phase, as well as in the magnitudes of the individual components (the latter is often referred to as an interaction-induced polarization shift). Although these results were already established in \cite{[4],APT04b,13}, the Hirota's formalism makes the calculation of the long-time asymptotics straightforward in both directions. An example of FS-FS solution is provided in Figs.~\ref{fig:fs-fs-interaction} for some specific choices of parameters. 

\subsection{Fundamental soliton - fundamental breather (FS-FB)}
\begin{figure}[H]
		\centering
		\includegraphics[width=0.75\linewidth]{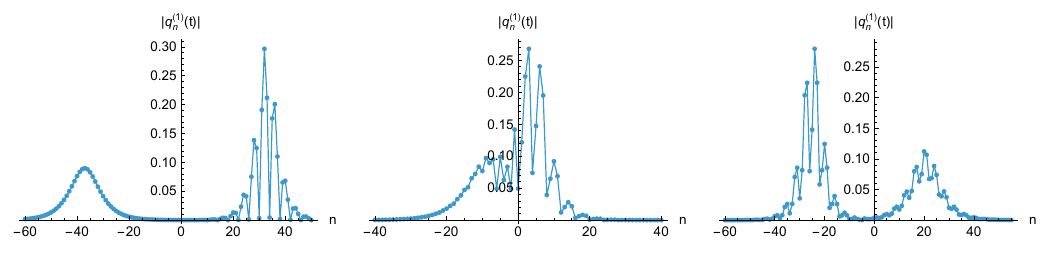}
		\includegraphics[width=0.75\linewidth]{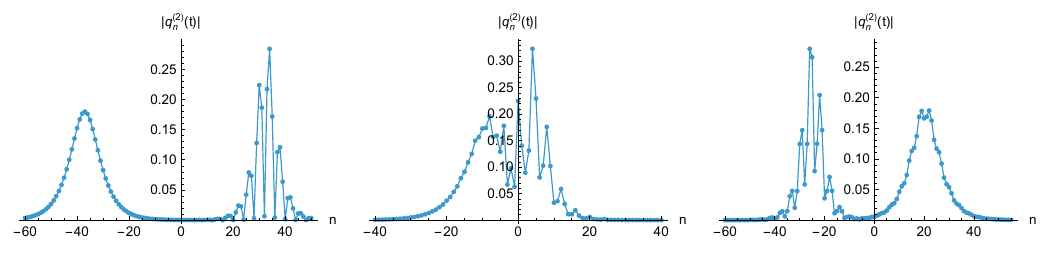}
		\caption{Snapshots of fundamental soliton-fundamental breather interaction at $t= -20,0, 20.$ The fundamental breather has parameters $p_1 = 3/10 + i \pi/4 $, $\bd \gamma = (1/10, 1/10)^T$, $\bd \mu = (1, 1)$ and fundamental soliton $p_2 =  1/5 -i \pi/4$, $\bd \kappa = (1,2)^T$.}
        \label{fig:fs-fb-snapshot}
	\end{figure}

\begin{figure}[H]
        \begin{subfigure}[t]{.45\textwidth}
        \centering
        \includegraphics[width=\linewidth]{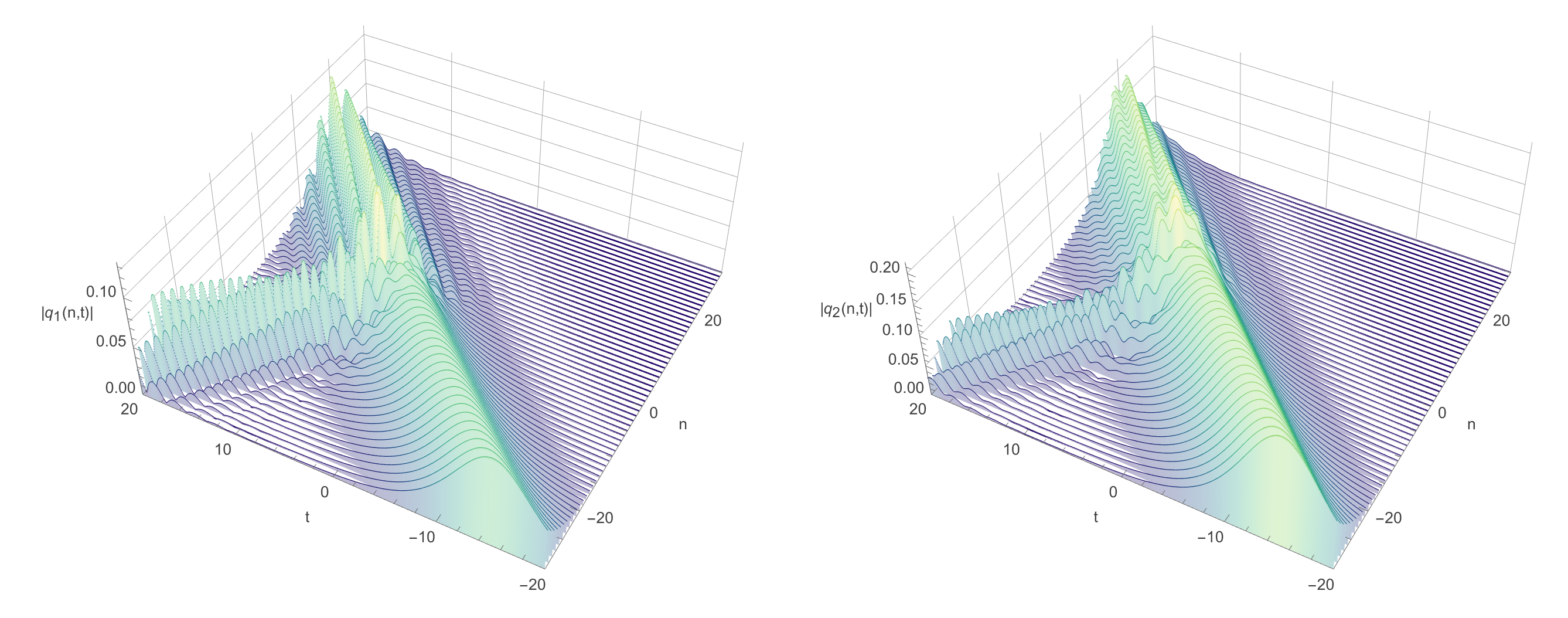}
        \caption{FS-FB solution with the FB asymptotics subtracted out in the $t\to -\infty$ direction}
        \end{subfigure}
    \hfill
        \begin{subfigure}[t]{.45\textwidth}
        \centering
        \includegraphics[width=\linewidth]{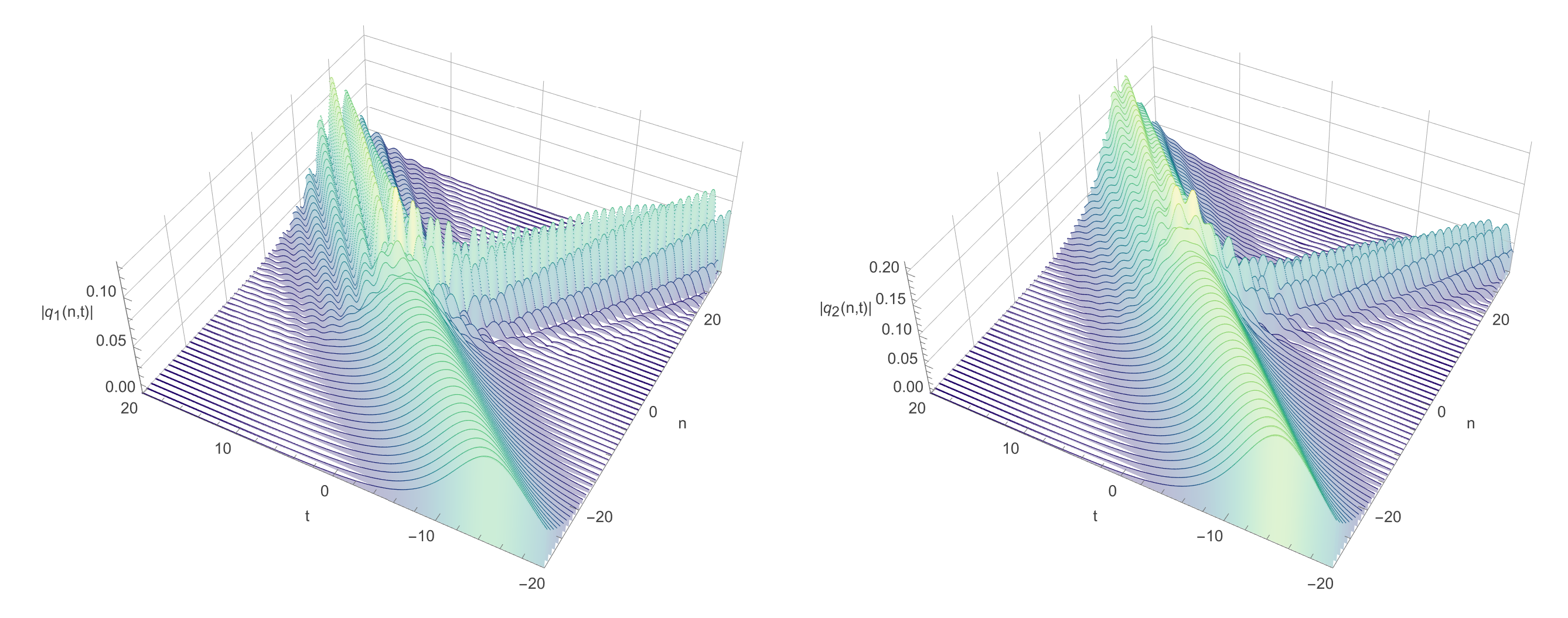}
        \caption{FS-FB solution with the FB asymptotics subtracted in the $t\to \infty$ direction}
        \end{subfigure}
    
        \begin{subfigure}[t]{.45\textwidth}
        \centering
        \includegraphics[width=\linewidth]{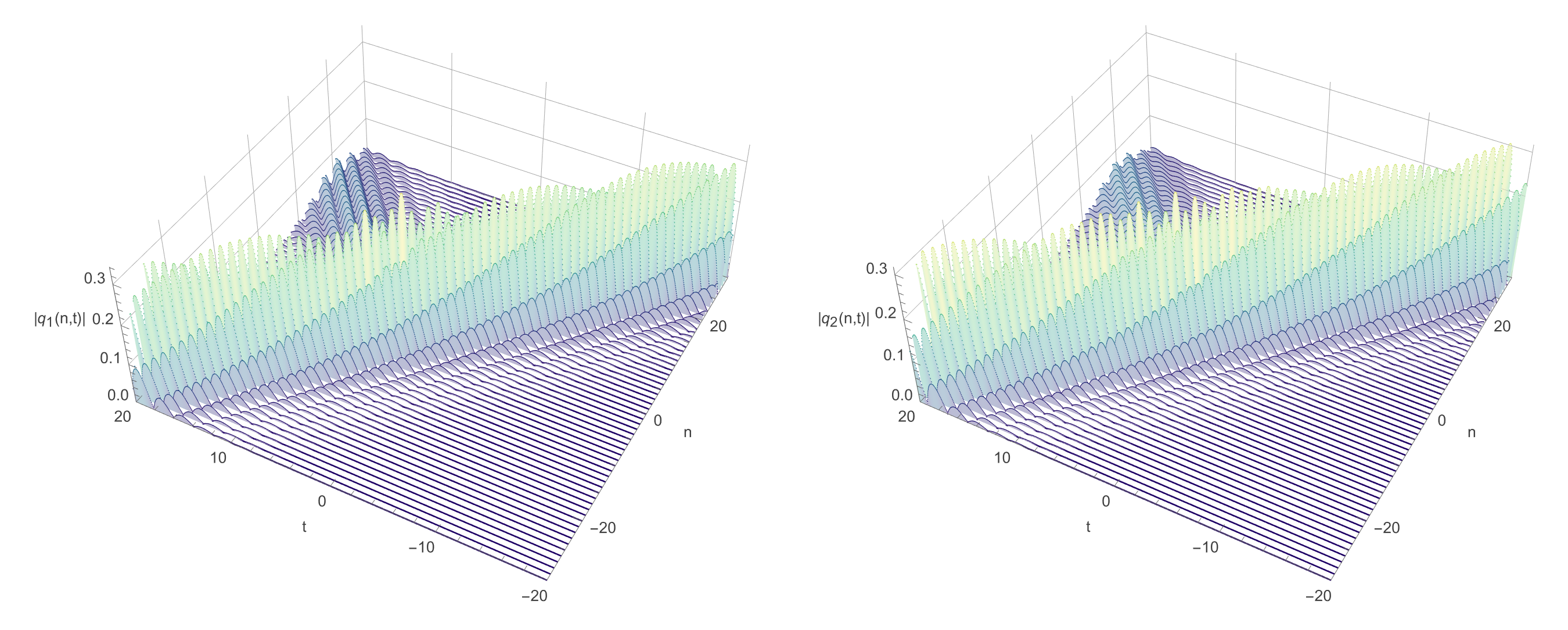}
        \caption{FS-FB solution with the FS asymptotics subtracted in the $t\to -\infty$ direction}
        \end{subfigure}
    \hfill
        \begin{subfigure}[t]{.45\textwidth}
        \centering
        \includegraphics[width=\linewidth]{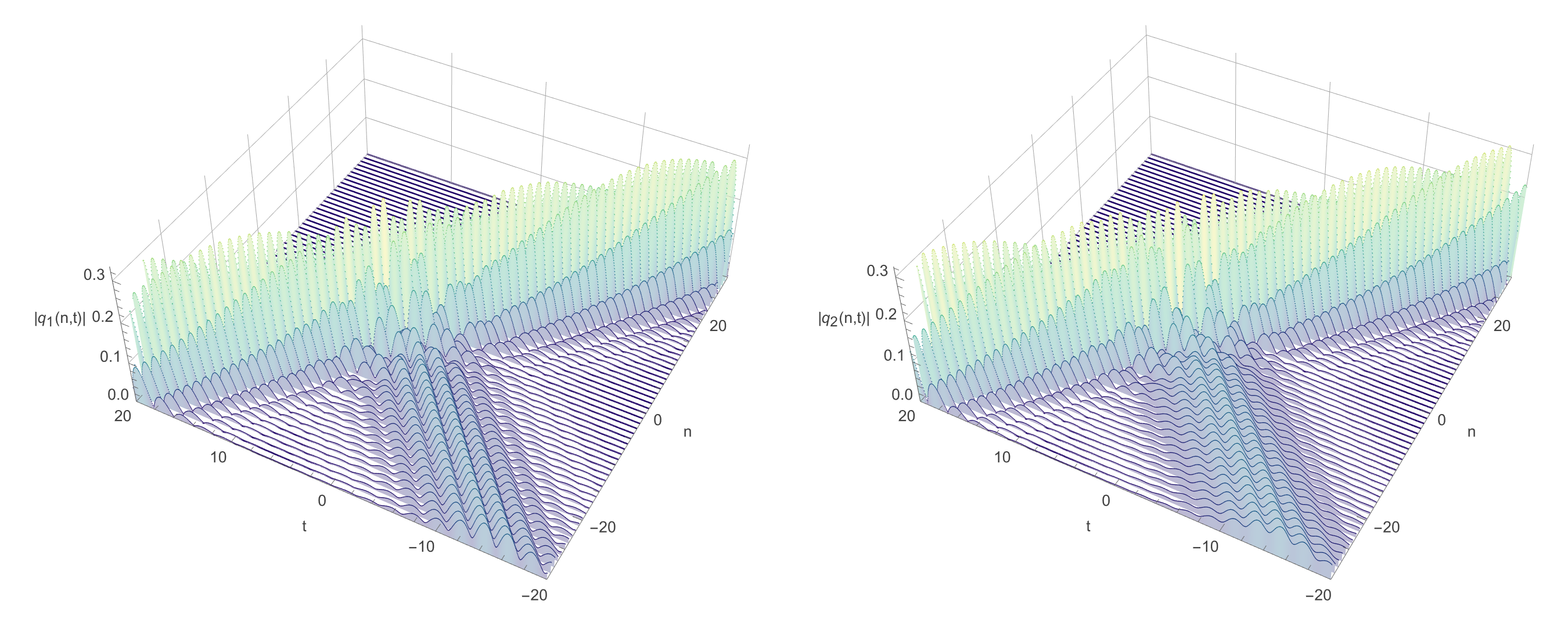}
        \caption{FS-FB solution with the FS asymptotics subtracted in the $t\to \infty$ direction}
        \end{subfigure}
    \caption{The same FS-FB solution as in Fig.~\ref{fig:fs-fb-snapshot}, but with the long-time asymptotics subtracted out.}
    \label{FS-FB asymptotics}
    \end{figure}

Here and in the subsequent sections, we will recycle the Greek letters used to denote the coefficients, while maintaining consistency within each subsection.  Let $E_1(n,t)$ and $\widetilde{E}_1(n,t)$ be the exponential functions associated with the fundamental breather and $E_2(n,t)$ the fundamental soliton.
The corresponding norming constant matrices are 
	\begin{align}
			\bd C_1 =  \left(\begin{array}{cc}
			\mu_1\gamma_1 & \mu_2\gamma_1\\ \mu_1\gamma_2 & \mu_2\gamma_2
		\end{array}\right), \quad \bd C_2 =  \left(\begin{array}{cc}
			\kappa_1 & 0\\ \kappa_2 & 0
		\end{array}\right),\quad  \mu_j, \gamma_j, \kappa_j \in \C, \quad j \in \{1,2\}.
	\end{align}
Then, the exact FS-FB solution is given by:
	\begin{equation}
		\bd q_n(t) = \left(\begin{array}{c}
			q^{(1)}_n(t)\\ q^{(2)}_n(t)\end{array}\right) = \frac{1}{1+ f_{n,2}+ f_{n,4}}\left(\begin{array}{c}
			g_{n,1}+ g_{n,3}\\ h_{n,1}+h_{n,3}
		\end{array}\right),
	\end{equation}
where
	\begin{subequations}\label{def FS-FB}
		\begin{gather}
			g_{n,1} = \mu_1^*\gamma_1^{*}E_1 + \mu_2(-\gamma_2)\widetilde{E}_1  + \kappa_1^{*}E_2,\quad 
			h_{n,1} = \mu_1^*\gamma_2^{*}E_1 + \mu_2\gamma_1\widetilde{E}_1  + \kappa_2^{*}E_2,\\
			f_{n,2} = \alpha_1 E_1 E_1^{*}  + \alpha_2 \widetilde{E}_1\widetilde{E}_1^{*} + \alpha_3 E_2E_2^{*}+ \alpha_4 E_1E_2^{*}+ \alpha_5E_2E_1^* + \alpha_6 E_2\widetilde{E}_1^*+ \alpha_7 \widetilde{E}_1E_2^*,\\
			g_{n,3} = \beta_1 E_1E_1^{*}E_2 +  \beta_2 \widetilde{E}_1\widetilde{E}_1^*E_2 + \beta_3 E_1E_2E_2^* + \beta_4 \widetilde{E}_1E_2E_2^* + \beta_5E_1\widetilde{E}_1E_2^* ,\\
			h_{n,3} = \delta_1 E_1E_1^{*}E_2 +  \delta_2 \widetilde{E}_1\widetilde{E}_1^*E_2 + \delta_3 E_1E_2E_2^* + \delta_4 \widetilde{E}_1E_2E_2^* + \delta_5E_1\widetilde{E}_1E_2^* ,\\
			f_{n,4} = \rho_1  E_1E_1^{*}E_2E_2^{*} +\rho_2 \widetilde{E}_1\widetilde{E}_1^{*}E_2E_2^{*},
		\end{gather}
	\end{subequations}
and
\begin{subequations}   
\begin{gather}
		\bd \gamma = \left(\begin{array}{c}
			\gamma_1 \\ \gamma_2 
		\end{array}\right),\quad  \bd \gamma^{\perp} = \left(\begin{array}{c}
		-\gamma_2 \\ \gamma_1 
	\end{array}\right),\quad \bd \kappa = \left(\begin{array}{c}
			\kappa_1 \\ \kappa_2 \end{array}\right),\\
		\alpha_1 = \langle \mu_1^*\bd\gamma^*, \mu_1 \bd\gamma \rangle A_{(1,1^*)},\quad \alpha_2 =  \langle\ \mu_2^*\bd\gamma^{\perp,*}, \mu_2\bd \gamma^{\perp}  \rangle A_{(\widetilde{1}, \widetilde{1}^*)},\quad \alpha_3 =  \langle \bd\kappa^* ,\bd \kappa\rangle A_{(2,2^*)},\\
        \alpha_4 =  \langle \mu_1^*\bd \gamma^*, \bd\kappa \rangle A_{(1,2^*)},\quad 
		\alpha_5 = \langle\bd \kappa^*, \mu_1\bd\gamma \rangle A_{(2,1^*)},\quad \alpha_6 =  \langle\bd \kappa^*, \mu_2\bd\gamma^{\perp,*} \rangle A_{(2, \widetilde{1}^*)},\\
        \alpha_7 =  \langle \mu_2\bd \gamma^{\perp}, \bd\kappa \rangle A_{(\widetilde{1},2^*)},\\
		\nonumber\\
		\beta_1 = P_{(1,2)}P_{(1, 1^*)}\alpha_5\mu_1^*\gamma_1^{*} -
		P_{(1,2)}P_{(2,1^*)}\alpha_1\kappa_1^*, \quad 
		\beta_2 = P_{(\widetilde{1},2)}P_{(\widetilde{1},\widetilde{1}^*)}\alpha_6\mu_2(-\gamma_2) - P_{(\widetilde{1},2)}P_{(2,\widetilde{1}^*)}\alpha_2\kappa_1^*,\\
		\beta_3 = P_{(1,2)}P_{(1,2^*)}\alpha_3\mu_1^*\gamma_1^* - P_{(1,2)}P_{(2,2^*)}\alpha_4\kappa_1^*,\quad
		\beta_4 = P_{(\widetilde{1},2)}P_{(\widetilde{1},2^*)}\alpha_3\mu_2(-\gamma_2) - P_{(\widetilde{1},2)}P_{(2,2^*)}\alpha_7\kappa_1^*,\\
		\beta_5 = P_{(1,\widetilde{1})}P_{(1,2^*)}\alpha_7\mu_1^*\gamma_1^* - P_{(1,\widetilde{1})}P_{(\widetilde{1},2^*)}\alpha_4\mu_2(-\gamma_2),\\
		\nonumber\\
		\delta_1 = P_{(1,2)}P_{(1, 1^*)}\alpha_5\mu_1^*\gamma_2^{*} -
		P_{(1,2)}P_{(2,1^*)}\alpha_1\kappa_2^*,\quad \delta_2 =P_{(\widetilde{1},2)}P_{(\widetilde{1},\widetilde{1}^*)}\alpha_6\mu_2\gamma_1 - P_{(\widetilde{1},2)}P_{(2,\widetilde{1}^*)}\alpha_2\kappa_2^*,\\
		\delta_3 = P_{(1,2)}P_{(1,2^*)}\alpha_3\mu_1^*\gamma_2^* - P_{(1,2)}P_{(2,2^*)}\alpha_4\kappa_2^*,\quad
		\delta_4 =P_{(\widetilde{1},2)}P_{(\widetilde{1},2^*)}\alpha_3\mu_2\gamma_1 - P_{(\widetilde{1},2)}P_{(2,2^*)}\alpha_7\kappa_2^*,\\
		\delta_5 = P_{(1,\widetilde{1})}P_{(1,2^*)}\alpha_7\mu_1^*\gamma_2^* - P_{(1,\widetilde{1})}P_{(\widetilde{1},2^*)}\alpha_4\mu_2\gamma_1,
		\end{gather}
		\begin{gather}
		\rho_1 = A_{(1,1^*,2,2^*)}\left(\frac{}{}\kappa_1^*\beta_1^*+ \kappa_2^* \delta_1^*+ \mu_1^*\gamma_1^*\beta_3^*  +\mu_1^*\gamma_2^*\delta_3^* + \kappa_1\beta_1+ \kappa_2 \delta_1+ \mu_1\gamma_1\beta_3  +\mu_1\gamma_2\delta_3 \right)\qquad\,\, \nonumber\\
        \qquad\qquad\qquad\qquad\qquad\qquad\qquad\qquad\qquad\qquad\qquad\qquad
		- S^2_{[(1,1^*),(2,2^*)]}\alpha_1\alpha_3- S^2_{[(1,2^*),(2,1^*)]}\alpha_4\alpha_5,\\ 
		\rho_2 = A_{(\widetilde{1},\widetilde{1}^*,2,2^*)}\left(\frac{}{} \kappa_1^*\beta_2^* + \kappa_2^*\delta_2^* + (-\mu_2\gamma_2) \beta_4^*  + \mu_2\gamma_1\delta_4^* +\kappa_1\beta_2 + \kappa_2\delta_2 + (-\mu_2^*\gamma_2^*) \beta_4  + \mu_2^*\gamma_1^*\delta_4  \right)\qquad\,\, \nonumber\\
        \qquad\qquad\qquad\qquad\qquad\qquad\qquad\qquad\qquad\qquad\qquad\qquad
			 - S^2_{[(\widetilde{1},\widetilde{1}^*),(2,2^*)]}\alpha_2\alpha_3- S^2_{[(2,\widetilde{1}^*),(\widetilde{1},2^*)]}\alpha_6\alpha_7, \\
		A_{(j,k^*)} = \frac{e^{p_j+p_k^{*}} \Delta h^2}{\left(e^{p_j+p_k^{*}}-1\right)^2},\quad 
		P_{(j,k^*)} = \frac{e^{p_j} + e^{p_k^{*}}}{e^{p_j+p_k^{*}}-1},\quad P_{(j,k)} = \frac{e^{p_j} - e^{p_k}}{e^{p_j+p_k}+1},\quad j,k \in \{1,2\},\\
		S_{[(j,k^*),(l,m^*)]}= \frac{e^{p_j+ p_k^*}- e^{p_l+ p_m^*}}{e^{p_j+p_k^*+p_l+p_m^*}-1}, \quad j,k,l,m \in \{1,2\}.
	\end{gather}
    \end{subequations}
Snapshots of a FS-FB solution at different times are plotted in Fig.~\ref{fig:fs-fb-snapshot}. The explicit expression of the solution provided above allows for a relatively straightforward calculation of the long-time asymptotics (see below), which previous determinantal formulas obscured.  
\paragraph{Long-time asymptotics of a FS-FB solution.}
We first express the functions in Eqs.~\eqref{def FS-FB} in terms of the real and imaginary parts of $E_j$ and $\widetilde{E}_j$, while noting that $\textrm{Re}(E_j)= \textrm{Re}(\widetilde{E}_j)$:
	\begin{subequations}
		\begin{align}
			g_{n,1} &= \mu_1^*\gamma_1^{*}e^{\xi_1 + i\nu_1} + \mu_2(-\gamma_2)e^{\xi_1 + i\widetilde{\nu}_1}  + \kappa_1^{*}e^{\xi_2 + i\nu_2},\\
			h_{n,1} &= \mu_1^*\gamma_2^{*}e^{\xi_1 + i\nu_1} + \mu_2\gamma_1e^{\xi_1 + i\widetilde{\nu}_1}  + \kappa_2^{*}e^{\xi_2 + i\nu_2},\\
			f_{n,2} &= \left( \alpha_1 + \alpha_2 \right)  e^{2\xi_1} + \alpha_3  e^{2\xi_2} + \alpha_4 e^{\xi_1+ \xi_2 + i (\nu_1-\nu_2)} + \alpha_5 e^{\xi_1+ \xi_2 + i(\nu_2-\nu_1)} \\
			 &\qquad + \alpha_6 e^{\xi_1+ \xi_2 + i(\nu_2 -\widetilde{\nu}_1)}+ \alpha_7 e^{\xi_1+ \xi_2 + i(\widetilde{\nu}_1-\nu_2)},\\
			g_{n,3} &= \left( \beta_1 + \beta_2\right) e^{2\xi_1+ \xi_2 + i\nu_2} + \beta_3 e^{2\xi_2 + \xi_1 + i\nu_1} + \beta_4 e^{2\xi_2 + \xi_1 + i\widetilde{\nu}_1}  \\
			& \qquad + \beta_5 e^{2\xi_1 + \xi_2 + i\left(\nu_1+ \widetilde{\nu}_1-\nu_2\right)},\\
			h_{n,3} &= \left(\delta_1 + \delta_2 \right) e^{2\xi_1+ \xi_2 + i\nu_2} + \delta_3 e^{2\xi_2 + \xi_1 + i\nu_1} + \delta_4 e^{2\xi_2 + \xi_1 + i\widetilde{\nu}_1}  \\
			& \qquad + \delta_5 e^{2\xi_1 + \xi_2 + i\left(\nu_1+ \widetilde{\nu}_1-\nu_2\right)}, \nonumber \\
			f_{n,4} &= \left(\rho_1  +\rho_2 \right) e^{2\xi_1 + 2\xi_2}.
		\end{align}
	\end{subequations}
Similar calculations as in the FS-FS case can be performed, and assuming that $v_2>v_1$, it can be seen that in the reference frame of the fundamental breather, i.e., for fixed $\xi_1$, if $t \to -\infty $, then $e^{\xi_2}\to \infty$ (cf. Eqs.~\eqref{e:xij},  \eqref{e:xi2vsxi1}, \eqref{e:xi1vsxi2}). The leading terms of $\bd q(n,t)$ are given by:
\begin{subequations}
\begin{gather}
		q^{(1)}_n(t)\sim \frac{\beta_3e^{\xi_1+ i\nu_1}+\beta_4 e^{\xi_1+ i\widetilde{\nu}_1}}{ \alpha_3  + \left(\rho_1 + \rho_2\right)e^{2\xi_1}},\quad 
		q^{(2)}_n(t)\sim \frac{\delta_3e^{\xi_1+ i\nu_1}+\delta_4 e^{\xi_1+ i\widetilde{\nu}_1}}{ \alpha_3  + \left(\rho_1 + \rho_2\right)e^{2\xi_1}}.
	\end{gather}
If $t\to \infty$, then $e^{\xi_2}\to 0$. The leading terms of $\bd q(n,t)$ are
	\begin{gather}
		q^{(1)}_n(t)\sim \frac{\mu_1^*\gamma_1^{*}e^{\xi_1 + i\nu_1} + \mu_2(-\gamma_2)e^{\xi_1 + i\widetilde{\nu}_1}}{1 + \left(\alpha_1 + \alpha_2\right)e^{2\xi_1} },\quad 
		q^{(2)}_n(t) \sim \frac{\mu_1^*\gamma_2^{*}e^{\xi_1 + i\nu_1} + \mu_2\gamma_1e^{\xi_1 + i\widetilde{\nu}_1}}{1 + \left(\alpha_1 + \alpha_2\right)e^{2\xi_1} }.
\end{gather}
\end{subequations}
Conversely, in the reference frame of the fundamental soliton, i.e., for fixed $\xi_2$, since $v_2 > v_1$ one has:
\begin{subequations}
\begin{alignat}{2}
		&t\to -\infty \implies e^{\xi_1}\to 0\colon \quad\quad&& q^{(1)}_n(t) \sim \frac{\kappa_1^{*}e^{\xi_2+ i\nu_2}}{1+\alpha_3 e^{2\xi_2}},\\
		& \quad\quad&& q^{(2)}_n(t) \sim \frac{\kappa_2^{*}e^{\xi_2+ i\nu_2}}{1+\alpha_3 e^{2\xi_2}},\\
		&t\to \infty \implies e^{\xi_1}\to \infty\colon \quad\quad&& q^{(1)}_n(t) \sim \frac{\left( \beta_1+ \beta_2\right)e^{\xi_2+ i \nu_2}+ \beta_5 e^{\xi_2 + i \left(\nu_1+ \widetilde{\nu}_1 - \nu_2 \right)}}{ \alpha_1 +\alpha_2 + \left(\rho_1+\rho_2\right)e^{\xi_2}},\\
		& \quad \quad&& q^{(2)}_n(t)\sim  \frac{\left( \delta_1+ \delta_2\right)e^{\xi_2+ i \nu_2}+ \delta_5 e^{\xi_2 + i \left(\nu_1+ \widetilde{\nu}_1 - \nu_2 \right)}}{ \alpha_1 +\alpha_2 + \left(\rho_1+\rho_2\right)e^{\xi_2}}.
\end{alignat}
\end{subequations}
Notice that from the definition of $p_j$, $\widetilde{p}_j$ it follows that:
	 \begin{align*}
	 	\nu_j &= 2nb_j + 2t - 2t \cosh(2a_j)\cos(2b_j),\\
	 	 \widetilde{\nu}_j &= n\pi- 2nb_j + 2t + 2t \cosh(2a_j)\cos(2b_j),
	 \end{align*}
hence
	\begin{align}
	 	\nu_1+ \widetilde{\nu}_1 - \nu_2 &=  2nb_1 + 2t - 2t \cosh(2a_1)\cos(2b_1) + n\pi- 2nb_1 + 2t + 2t \cosh(2a_1)\cos(2b_1) \nonumber \\
	 	& \quad - 2n b_2 -2t + 2t\cosh(2a_2)\cos(2b_2), \nonumber\\
	 	&= n\pi - 2nb_2 + 2t + 2t\cosh(2a_2)\cos(2b_2),\nonumber\\
	 	&= \widetilde{\nu}_2.
	 \end{align}
The above asymptotics as $t\to \infty$ proves that the fundamental soliton emerges as a fundamental breather after the interaction (as can be also seen from the snapshots in Fig.~\ref{fig:fs-fb-snapshot}), while also providing the explicit expressions of the FS and FB after the interaction. The correctness of the long-time asymptotics can be verified numerically by subtracting the predicted asymptotic behavior in each direction from the exact FS-FB solution, as shown in Fig.~\ref{FS-FB asymptotics}.
    
\subsection{Fundamental breather - fundamental breather (FB-FB)}
Let $E_1(n,t)$ and $\widetilde{E}_1(n,t)$ be the exponential functions associated with the first fundamental breather and $E_2(n,t)$, $\widetilde{E}_2(n,t)$ the second fundamental breather. The corresponding norming constant matrices are 
	\begin{align}
		\bd C_1 =  \left(\begin{array}{cc}
			\mu_1\gamma_1 & \mu_2\gamma_1\\ \mu_1\gamma_2 & \mu_2\gamma_2
		\end{array}\right), \quad \bd C_2 =  \left(\begin{array}{cc}
			\omega_1\kappa_1 & \omega_2\kappa_1\\ \omega_1\kappa_2 & \omega_2 \kappa_2
		\end{array}\right),\quad  \mu_j,\omega_j, \gamma_j, \kappa_j \in \C, \quad j \in \{1,2\}.
	\end{align}
	Then, the exact FB-FB solution is given by: 
	\begin{equation}
  \label{e:FBFB}  
		\bd q_n(t) = \left(\begin{array}{c}
			q^{(1)}_n(t)\\ q^{(2)}_n(t)\end{array}\right) = \frac{1}{1+ f_{n,2}+ f_{n,4}}\left(\begin{array}{c}
			g_{n,1}+ g_{n,3}\\ h_{n,1} + h_{n,3}
		\end{array}\right),
	\end{equation}
where
\begin{subequations}
\label{e:FBFBexp}
		\begin{gather}
			g_{n,1} = \mu_1^*\gamma_1^{*}E_1 + (-\mu_2\gamma_2)\widetilde{E}_1+ \omega_1^*\kappa_1^{*}E_2 + (-\omega_2\kappa_2) \widetilde{E}_2,\\
			h_{n,1} = \mu_1^*\gamma_2^{*}E_1 + \mu_2\gamma_1\widetilde{E}_1+ \omega_1^*\kappa_2^{*}E_2 + \omega_2\kappa_1 \widetilde{E}_2,\\
			f_{n,2} = \alpha_1 E_1 E_1^{*}  + \alpha_2 E_1E_2^{*} + \alpha_3 E_1\widetilde{E}_2^{*}+ \alpha_4 E_2E_1^{*}+ \alpha_5  E_2E_2^{*} + \alpha_6  E_2\widetilde{E}_1^{*} \qquad\qquad\quad \\ + \alpha_7 \widetilde{E}_1E_2^{*} + \alpha_8  \widetilde{E}_1\widetilde{E}_1^{*}
            + \alpha_9  \widetilde{E}_1\widetilde{E}_2^{*} + \alpha_{10}  \widetilde{E}_2E_1^{*} +\alpha_{11}  \widetilde{E}_2\widetilde{E}_1^{*} + \alpha_{12}  \widetilde{E}_2\widetilde{E}_2^{*}, \nonumber \\
			g_{n,3} = \beta_1 E_1E_1^{*}E_2 +  \beta_2 E_1E_2E_2^* +\beta_3 E_1\widetilde{E}_1E_2^{*} +\beta_4 E_1\widetilde{E}_1\widetilde{E}_2^{*} +\beta_5 \widetilde{E}_1E_2E_2^{*} + \beta_6 \widetilde{E}_1\widetilde{E}_1^{*}E_2 \qquad\quad\,\,\,\\
            +\beta_7 E_1E_1^{*}\widetilde{E}_2 +\beta_8E_1 \widetilde{E}_2\widetilde{E}_2^{*} +\beta_9 E_2\widetilde{E}_2E_1^{*} + \beta_{10}E_2\widetilde{E}_2\widetilde{E}_1^{*} +\beta_{11} \widetilde{E}_1\widetilde{E}_1^{*}\widetilde{E}_2 +\beta_{12} \widetilde{E}_2\widetilde{E}_2^{*}\widetilde{E}_1, \nonumber \\
			h_{n,3} = \delta_1 E_1E_1^{*}E_2 +  \delta_2 E_1E_2E_2^* +\delta_3 E_1\widetilde{E}_1E_2^{*} +\delta_4 E_1\widetilde{E}_1\widetilde{E}_2^{*} + \delta_5 \widetilde{E}_1E_2E_2^{*} + \delta_6 \widetilde{E}_1\widetilde{E}_1^{*}E_2\qquad\qquad\\
            +\delta_7 E_1E_1^{*}\widetilde{E}_2 +\delta_8 E_1 \widetilde{E}_2\widetilde{E}_2^{*}
			+\delta_9 E_2\widetilde{E}_2E_1^{*} + \delta_{10}E_2\widetilde{E}_2\widetilde{E}_1^{*} +\delta_{11} \widetilde{E}_1\widetilde{E}_1^{*}\widetilde{E}_2 +\delta_{12} \widetilde{E}_2\widetilde{E}_2^{*}\widetilde{E}_1, \nonumber \\
			f_{n,4} = \rho_1  E_1E_1^{*}E_2E_2^{*} + \rho_2 \widetilde{E}_1\widetilde{E}_1^{*}E_2E_2^{*} +\rho_3 E_1E_1^{*}\widetilde{E}_2\widetilde{E}_2^{*} \qquad\qquad\qquad\\ +\rho_4 \widetilde{E}_1\widetilde{E}_1^{*}\widetilde{E}_2\widetilde{E}_2^{*} + \rho_5 E_1\widetilde{E}_1E_2^{*}\widetilde{E}_2^{*} +\rho_6 E_1^{*}\widetilde{E}_1^{*}E_2\widetilde{E}_2, \nonumber
		\end{gather}
	\end{subequations}
and {the coefficients of $g_{n,3}$, $h_{n,3}$, $f_{n,2}$, and $f_{n,4}$ are defined in Appendix \ref{ssec: FB-FB coefs}}. Fig.~\ref{fig:fb-fb-interaction} shows some snapshots of a FB-FB solution for specific choices of the breather parameters, while its long-time asymptotic behavior is computed below.
    \begin{figure}[H]
		\centering
		\includegraphics[width=0.7\linewidth]{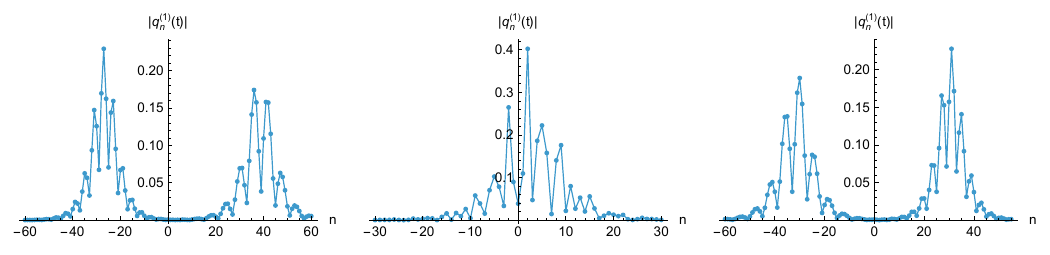}
		\includegraphics[width=0.7\linewidth]{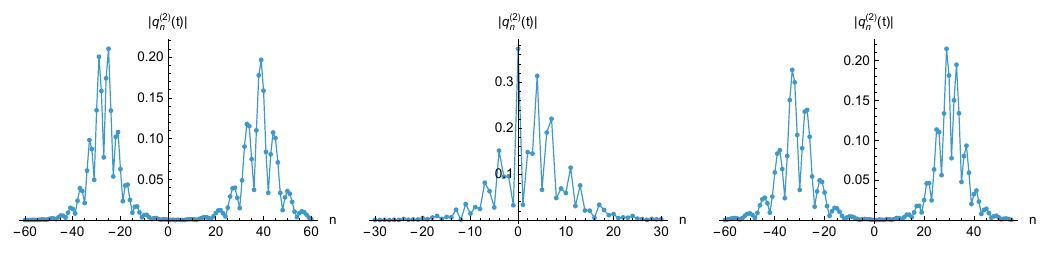}
		\caption{Snapshots of the fundamental breather-fundamental breather interaction with $p_1 = 1/5+ i \left(2\pi/3\right) $, $\bd \gamma = (2,3)^T$, $\bd \mu = (1/25, 1/25)^T$, $p_2 = 6/25 -i\pi/4 $, $\bd \kappa = (1/5,1/5)^T$, $\bd \omega = (1, 1/2)^T$ at $t =  -20, 0 , 20.$}
        \label{fig:fb-fb-interaction}
	\end{figure}
\paragraph{Long-time asymptotics of the FB-FB solutions.}
	Similarly to previous cases, we express the solution $\bd q(n,t)$ in terms of the real and imaginary parts of $E_j$ and $\widetilde{E}_j$. In the reference frame of the first fundamental breather, { the one} associated with $\xi_1$, and { assuming}, as before, $v_2 > v_1$, we find:
\begin{subequations}
    \begin{alignat}{2}
		&t\to -\infty \implies e^{\xi_2}\to \infty\colon \quad\quad&& q^{(1)}_n(t) \sim \frac{\left(\beta_2+\beta_8 + \beta_{10}\right)e^{\xi_1+ i\nu_1} + \left(\beta_5 + \beta_9 + \beta_{12}\right)e^{\xi_1+ i\widetilde{\nu}_1}}{(\alpha_5+ \alpha_{12})+ \left(\sum_{k=1}^{6}\rho_k\right)e^{2\xi_1} } ,\\
		& \quad\quad&& q^{(2)}_n(t) \sim \frac{\left(\delta_2+\delta_8 + \delta_{10}\right)e^{\xi_1+ i\nu_1} + \left(\delta_5 + \delta_9 + \delta_{12}\right)e^{\xi_1+ i\widetilde{\nu}_1}}{(\alpha_5+ \alpha_{12})+ \left(\sum_{k=1}^{6}\rho_k\right)e^{2\xi_1} } ,\\
		&t\to \infty \implies e^{\xi_2}\to 0\colon \quad\quad&& q^{(1)}_n(t) \sim \frac{\mu_1^*\gamma_1^{*}e^{\xi_1+ i\nu_1} + (-\mu_2\gamma_2)e^{\xi_1+ i\widetilde{\nu}_1}}{1+ (\alpha_1 + \alpha_8)e^{2\xi_1}},\\
		& \quad \quad&& q^{(2)}_n(t)\sim  \frac{\mu_1^*\gamma_2^{*}e^{\xi_1+ i\nu_1} + \mu_2\gamma_1e^{\xi_1+ i\widetilde{\nu}_1}}{1+ (\alpha_1 + \alpha_8)e^{2\xi_1}}.
	\end{alignat}
\end{subequations}    
Analogously, in the reference frame of the second fundamental breather, i.e., for fixed $\xi_2$, we find:
\begin{subequations}
		\begin{alignat}{2}
			&t\to -\infty \implies e^{\xi_1}\to 0\colon \quad\quad&& q^{(1)}_n(t) \sim  \frac{\omega_1^*\kappa_1^{*}e^{\xi_2+ i\nu_2} + (-\omega_2\kappa_2) e^{\xi_2+ i\widetilde{\nu}_2}}{1+ \left(\alpha_5 + \alpha_{12}\right)e^{2\xi_2}},\\
			& \quad\quad&& q^{(2)}_n(t) \sim \frac{\omega_1^*\kappa_2^{*}e^{\xi_2+ i\nu_2} + \omega_2\kappa_1 e^{\xi_2+ i\widetilde{\nu}_2}}{1+ \left(\alpha_5 + \alpha_{12}\right)e^{2\xi_2}} ,\\
			&t\to \infty \implies e^{\xi_1}\to \infty\colon \quad\quad&& q^{(1)}_n(t) \sim \frac{\left( \beta_1 + \beta_4 + \beta_6\right)e^{\xi_2+i\nu_2}+ \left( \beta_3 + \beta_7 + \beta_{11}\right)e^{\xi_2+i\widetilde{\nu}_2}}{ \left(\alpha_1 + \alpha_8\right) + \left(\sum_{k=1}^{6}\rho_k\right)e^{2\xi_2} } ,\\
			& \quad \quad&& q^{(2)}_n(t)\sim  \frac{\left( \delta_1 + \delta_4 + \delta_6\right)e^{\xi_2+i\nu_2}+ \left( \delta_3 + \delta_7 + \delta_{11}\right)e^{\xi_2+i\widetilde{\nu}_2}}{ \left(\alpha_1 + \alpha_8\right) + \left(\sum_{k=1}^{6}\rho_k\right)e^{2\xi_2} }.
		\end{alignat}
 \end{subequations} 
 Both FBs emerge from the interaction as such, consistently with the snapshots in Fig.~\ref{fig:fb-fb-interaction}, and the phase, center and polarization and shifts can be evaluated explicitly in terms of the coefficients in Appendix \ref{ssec: FB-FB coefs}. The validity of the above asymptotics has been verified numerically as done in Fig.~\ref{FS-FB asymptotics}, although the corresponding plots have been omitted for brevity.
 
	\subsection{Fundamental soliton - composite breather (FS-CB)}
Let the functions $E_1(n,t)$ and $\widetilde{E}_1(n,t)$ be the exponential functions associated with the CB, and $E_2(n,t)$ the one associated with the FS. The corresponding norming constant matrices are given as:
	\begin{align}
		\bd C_1 =  \left(\begin{array}{cc}
			\gamma_1 & \epsilon_1 \\ \gamma_2 & \epsilon_2
		\end{array}\right), \quad \bd C_2 =  \left(\begin{array}{cc}
			\kappa_1 & 0\\ \kappa_2 & 0
		\end{array}\right),\quad  \epsilon_j, \gamma_j, \kappa_j \in \C, \quad j \in \{1,2\}.
	\end{align}
Starting { the} Hirota's perturbation scheme with 
	\begin{align*}
		g_{n,1} &= \gamma_1^{*}E_1(n,t)+ \epsilon_1\widetilde{E}_1(n,t) + \kappa_1^{*}E_2(n,t) ,\\
		h_{n,1} &=\gamma_2^{*}E_1(n,t)+ \epsilon_2\widetilde{E}_1(n,t) + \kappa_2^{*}E_2(n,t),
	\end{align*} 
yields the follwoing expression for the FB-CB solution:
	\begin{equation}\label{FS-CB}
		\bd q_n(t) = \left(\begin{array}{c}
			q^{(1)}_n(t)\\ q^{(2)}_n(t)\end{array}\right) = \frac{1}{1+ \sum_{k=1}^{3}f_{n,2k}}\left(\begin{array}{c}
			\sum_{k=1}^{3}g_{n, 2k-1}\\ \sum_{k=1}^{3}h_{n, 2k-1}
		\end{array}\right),
	\end{equation} 
with functions $g_{n,2k-1}$, $h_{n,2k-1}$ and $f_{n,2k}$ for $k= 1,2,3$ given by:
	 \begin{gather*}
	 	g_{n,1} = \gamma_1^{*}E_1 + \epsilon_1\widetilde{E}_1 + \kappa_1^{*}E_2,\\
	 	g_{n,3} = \beta_1 E_1E_1^{*}\widetilde{E}_1 + \beta_2E_1E_1^{*}E_2 +\beta_3 E_1 \widetilde{E}_1\widetilde{E}_1^{*} + \beta_4 E_1\widetilde{E}_1^{*}E_2 + \beta_5 E_1\widetilde{E}_1E_2^{*} \\+ \beta_6 E_1E_2E_2^{*} + \beta_7\widetilde{E}_1E_1^{*}E_2 
        + \beta_8 \widetilde{E}_1\widetilde{E}_1^{*}E_2 + \beta_9 \widetilde{E}_1E_2E_2^{*}, \\
	 	g_{n,5} = \mu_1 E_1E_1^{*}\widetilde{E}_1E_2E_2^{*}+ \mu_2 E_1E_1^{*}\widetilde{E}_1\widetilde{E}_1^{*}E_2+ \mu_3 E_1\widetilde{E}_1\widetilde{E}_1^{*}E_2E_2^{*}, \\
	 	\\
	 	h_{n,1} =  \gamma_2^{*}E_1(n,t) +\epsilon_2\widetilde{E}_1(n,t) + \kappa_2^{*} E_2(n,t),\\
	 	h_{n,3} = \delta_1 E_1E_1^{*}\widetilde{E}_1 + \delta_2E_1E_1^{*}E_2 +\delta_3 E_1 \widetilde{E}_1\widetilde{E}_1^{*} + \delta_4 E_1\widetilde{E}_1^{*}E_2 + \delta_5 E_1\widetilde{E}_1E_2^{*} \\ + \delta_6 E_1E_2E_2^{*} + \delta_7\widetilde{E}_1E_1^{*}E_2
        + \delta_8 \widetilde{E}_1\widetilde{E}_1^{*}E_2 + \delta_9 \widetilde{E}_1E_2E_2^{*}, \\
	 	h_{n,5} = \sigma_1 E_1E_1^{*}\widetilde{E}_1E_2E_2^{*}+ \sigma_2 E_1E_1^{*}\widetilde{E}_1\widetilde{E}_1^{*}E_2+ \sigma_3 E_1\widetilde{E}_1\widetilde{E}_1^{*}E_2E_2^{*},  \\
	 	\\
	 	f_{n,2} = \alpha_1 E_1E_1^{*} + \alpha_2 E_1\widetilde{E}_1^{*} + \alpha_3 E_1E_2^{*} + \alpha_4 \widetilde{E}_1E_1^{*} + \alpha_5 \widetilde{E}_1\widetilde{E}_1^{*} + \alpha_6 E_1^{*}E_2 + \alpha_7 E_2E_2^{*}, \\
	 	f_{n,4} = \rho_1 E_1E_1^{*}E_2E_2^{*}+ \rho_2 E_1E_1^{*}\widetilde{E}_1^{*}E_2+ \rho_3 E_1\widetilde{E}_1^{*}E_2E_2^{*}+ \rho_4E_1E_1^{*}\widetilde{E}_1E_2^{*}\\
        + \rho_5 E_1E_1^{*}\widetilde{E}_1\widetilde{E}_1^{*} + \rho_6 \widetilde{E}_1E_1^{*}E_2E_2^{*} + \rho_7 \widetilde{E}_1\widetilde{E}_1^{*}E_2E_2^{*}, \\
	 	f_{n,6} = \rho_8 E_1E_1^{*}\widetilde{E}_1\widetilde{E}_1^{*}E_2E_2^{*}.
	 \end{gather*}
Since the expressions of the coefficients of $g_{n,2k-1}$, $h_{n,2k-1}$ and $f_{n,2k}$ are very lengthy, and listing them has little practical use, for illustrative purposes we computed the coefficients for the following  parameters:	
\begin{gather}
\label{FS-CB_par}
		p_1  = \frac{2}{5}+ i \frac{\pi}{4},\quad  p_2 = \frac{2}{5} ,\quad 
		 \bd C_1 =  \left(\begin{array}{cc}
		 	1 & -\frac{1}{10} \\ 2 & \frac{1}{10}
		 \end{array}\right), \quad   \bd C_2 =  \left(\begin{array}{cc}
		 1 & 0\\ 1 & 0
	 \end{array}\right).
\end{gather}
The values of the coefficients for these parameter choices are provided in Appendix \ref{ssec: FS-CB coefs}.
Fig.~\ref{fig:CB-FS snapshots} shows snapshots of the FS-CB solution with parameters as in Eq.~\eqref{FS-CB_par}.
\begin{figure}[H]
		\centering
		\includegraphics[width=0.75\linewidth]{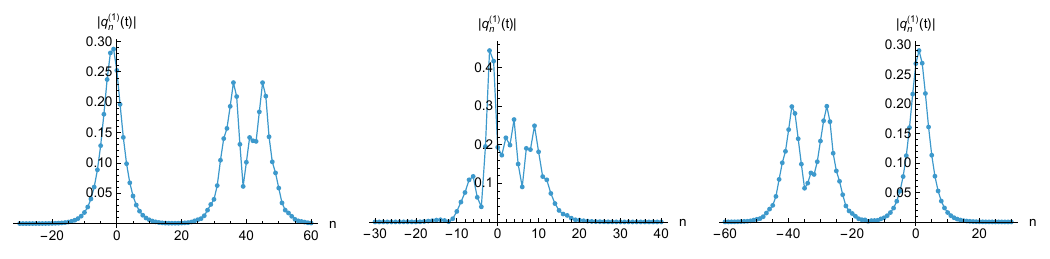}
		\includegraphics[width=0.75\linewidth]{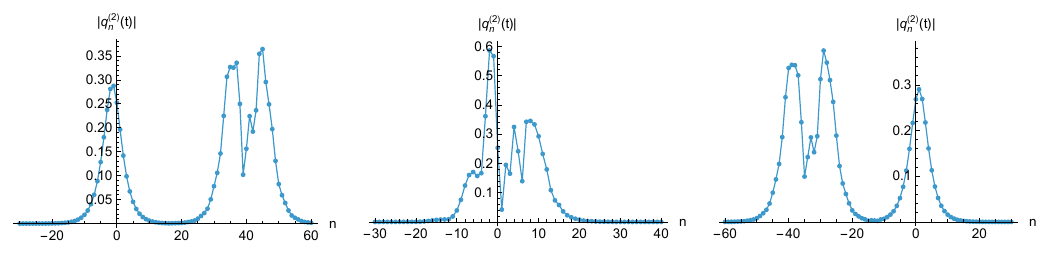}
		\caption{Snapshots at times $t= -25, 0$, and $25$ of the FS-CB solution with parameters as in Eq.~\eqref{FS-CB_par}.}
		\label{fig:CB-FS snapshots}
\end{figure}
\paragraph{Long-time asymptotics of the FS-CB solutions.}
Similarly to previous cases, the solution $\bd q_n(t)$ is expressed in terms of the real and imaginary part of $E_i$ and $\widetilde{E}_i$ for $i= 1,2$. In the reference frame of the CB, we fix $\xi_1$ and express $\xi_2$ as in Eq.~\eqref{e:xi2vsxi1}.
Assuming $v_2> v_1$, we find  $e^{\xi_2}\to \infty$ as $t\to -\infty$. Hence, the leading behavior of each function $g_j$, $h_j$ and $f_j$ is given by exponential terms which have the largest coefficients of $\xi_2$: 
    \small
    \begin{subequations}\label{FS-CB ref frame1 tpos}
	\begin{alignat}{3}
		&t \to -\infty \implies e^{\xi_2}\to \infty\colon \quad\quad  && q^{(1)}_n(t) &&\sim \frac{\beta_6 e^{\xi_1 + i\nu_1} + \beta_9 e^{ \xi_1 + i\widetilde{\nu}_1} + e^{2\xi_1}\left (\mu_1e^{ \xi_1 + i\widetilde{\nu}_1} +\mu_3 e^{\xi_1 + i\nu_1}\right)}{ \alpha_7 + e^{2\xi_1}\left(\left(\rho_1 + \rho_7 \right)+ \rho_3 e^{i(\nu_1- \widetilde{\nu}_1)} + \rho_6 e^{-i(\nu_1 - \widetilde{\nu}_1})  \right) + \rho_8 e^{4\xi_1} },\\
	& \quad && q^{(2)}_n(t) && \sim \frac{\delta_6 e^{\xi_1+ i\nu_1} + \delta_9 e^{\xi_1 + i\widetilde{\nu}_1} + e^{2\xi_1}\left(\sigma_1 e^{\xi_1 + i\widetilde{\nu}_1} + \sigma_3 e^{\xi_1+i\nu_1} \right)  }{ \alpha_7 + e^{2\xi_1}\left(\left(\rho_1 + \rho_7 \right)+ \rho_3 e^{i(\nu_1- \widetilde{\nu}_1)} + \rho_6 e^{-i(\nu_1 - \widetilde{\nu}_1})  \right) + \rho_8 e^{4\xi_1} }.
	\end{alignat}
	\end{subequations} 
    \normalsize
	In the limit $t \to \infty$, $e^{\xi_2}\to 0$.  Hence, exponential terms associated with the soliton can be neglected in this limit. Then, the leading behavior of $\bd q(n,t)$ is given by:
	\begin{subequations}\label{FS-CB ref frame1 tneg}
	\begin{alignat}{3}
		& t\to \infty \implies e^{\xi_2} \to 0\colon \quad &&q^{(1)}_n(t) &&\sim \frac{\gamma_1^{*} e^{\xi_1 + i\nu_1} + \epsilon_1 e^{ \xi_1 + i\widetilde{\nu}_1} + e^{2\xi_1}\left (\beta_1 e^{ \xi_1 + i\widetilde{\nu}_1} +\beta_3 e^{\xi_1 + i\nu_1}\right)}{ 1+ e^{2\xi_1}\left(\left(\alpha_1 + \alpha_5 \right)+ \alpha_2 e^{i(\nu_1- \widetilde{\nu}_1)} + \alpha_4 e^{-i(\nu_1 - \widetilde{\nu}_1})  \right) + \rho_5 e^{4\xi_1} },  \\
		&  \quad && q^{(2)}_n(t) &&\sim  \frac{\gamma_2^{*} e^{\xi_1 + i\nu_1} + \epsilon_2 e^{ \xi_1 + i\widetilde{\nu}_1} + e^{2\xi_1}\left (\delta_1 e^{ \xi_1 + i\widetilde{\nu}_1} +\delta_3 e^{\xi_1 + i\nu_1}\right)}{ 1+ e^{2\xi_1}\left(\left(\alpha_1 + \alpha_5 \right)+ \alpha_2 e^{i(\nu_1- \widetilde{\nu}_1)} + \alpha_4 e^{-i(\nu_1 - \widetilde{\nu}_1})  \right) + \rho_5 e^{4\xi_1} }.
	\end{alignat}
	\end{subequations}
Consistently with the snapshots in Fig.~\ref{fig:CB-FS snapshots} (and the predictions in \cite{13} via the Manakov method), Eqs.~\eqref{FS-CB ref frame1 tpos} and \eqref{FS-CB ref frame1 tneg} confirm that the CB remains a CB after the interaction with the FS, only acquiring a phase shift.

\begin{figure}[H]
		\centering
		\includegraphics[width=0.4\linewidth]{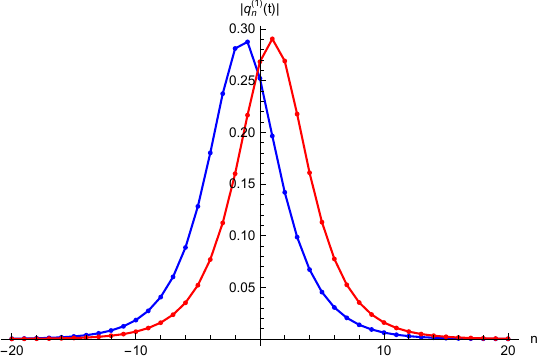}
		\caption{ Superimposition of the snapshots of the asymptotics of $q^{(1)}_n(t)$ as $t\to - \infty$ (blue) and $t\to \infty$ (red) in the reference frame of the soliton $\xi_2$. }
		\label{fig:fs-cb-shift-center-soliton}
	\end{figure}

Similarly, in the reference frame of the soliton, we fix $\xi_2$ and express $\xi_1$ via Eq.~\eqref{e:xi1vsxi2}.
Then, since $v_2>v_1$, as $t\to -\infty$, we have $e^{\xi_1}\to 0$. Hence, exponential terms associated with the breather can be neglected in this direction. The leading behavior of $q^{(1)}_n(t)$ and $q^{(2)}_n(t)$ is given by: 
    \begin{subequations}\label{FS-CB ref frame2 tneg}
	\begin{gather}
		q^{(1)}_n(t) \sim  \frac{\kappa_1^{*}e^{\xi_2+ i\nu_2}}{1+ \alpha_7 e^{2\xi_2}},\quad \quad 
		q^{(2)}_n(t) \sim \frac{\kappa_2^{*}e^{\xi_2+ i\nu_2}}{1+ \alpha_7e^{2\xi_2}}.
	\end{gather}
	\end{subequations}
For $t \to \infty$, $e^{\xi_1}\to \infty$ and therefore the asymptotic behavior of $\bd q_n(t)$ is determined by exponentials with the largest coefficients of $\xi_1$, namely:
	\begin{subequations}\label{FS-CB ref frame2 tpos}
	\begin{alignat}{2}
		q^{(1)}_n(t) &\sim 
        \frac{\mu_2 e^{\xi_2+ i\nu_2}}{\rho_5 + \rho_8 e^{2\xi_2}},\qquad q^{(2)}_n(t) 
			\sim \frac{\sigma_2 e^{\xi_2}}{\rho_5 + \rho_8 e^{2\xi_2}  }.
	\end{alignat}
	\end{subequations} 
Again, the long-time asymptotics of the soliton confirm that its form is preserved after interacting with the CB, although the center of the soliton is shifted, see Fig. \ref{fig:fs-cb-shift-center-soliton}.

	\subsection{Fundamental breather - composite breather (FB-CB)}
\begin{figure}[H]
		\centering
		\includegraphics[width=0.7\linewidth]{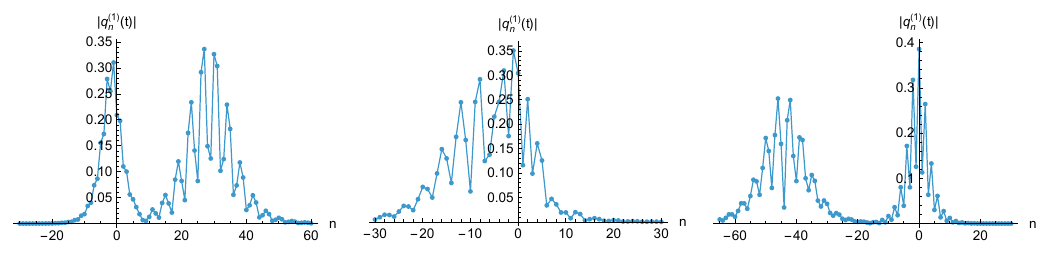}
		\includegraphics[width=0.7\linewidth]{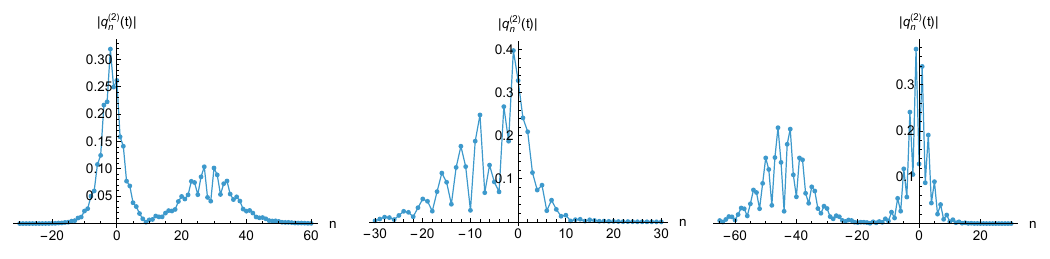}
		\caption{Snapshots at times $t =  -20, 0 , 20$  of a FB-CB solution for the parameters in Eqs.~\eqref{CB-FB par}. }
        \label{fig:CB-FB snapshots}
	\end{figure}

Let $E_1$ and $\widetilde{E}_1$ be the exponential functions associated with the composite breather, and $E_2$ and $\widetilde{E}_2$ to the fundamental breather. The corresponding norming constant matrices are given as:
	\begin{align}
		\bd C_1 =  \left(\begin{array}{cc}
			\gamma_1 & \epsilon_1 \\ \gamma_2 & \epsilon_2
		\end{array}\right), \quad \bd C_2 =  \left(\begin{array}{cc}
			\mu_1\kappa_1 & \mu_2\kappa_1 \\ \mu_1\kappa_2 & \mu_2\kappa_2
		\end{array}\right),\quad  \epsilon_j, \gamma_j, \kappa_j, \mu_j \in \C, \quad j \in \{1,2\}.
	\end{align}
Solving for { $g_{n,2k-1}, h_{n,2k-1}$ and $f_{n,2k}$} order by order in $\varepsilon$ in the perturbation like expansion starting with the ansatz
\begin{align*}
		g_{n,1} &= \gamma_1^{*} E_1 + \epsilon_1\widetilde{E}_1 + \mu_1^{*}\kappa_1^{*}E_2 + (-\mu_2)\kappa_2\widetilde{E}_2,\\
		h_{n,1} &= \gamma_2^{*} E_1 + \epsilon_2\widetilde{E}_1 + \mu_1^{*}\kappa_2^{*}E_2 + \mu_2\kappa_1\widetilde{E}_2,
\end{align*}
gives the FB-CB solution in the form:
\begin{equation}
		\bd q_n(t) = \left(\begin{array}{c}
			q^{(1)}_n(t)\\ q^{(2)}_n(t)\end{array}\right) = \frac{1}{1+ \sum_{k=1}^{3}f_{n,2k}}\left(\begin{array}{c}
			\sum_{k=1}^{3}g_{n,2k-1}\\ \sum_{k=1}^{3}h_{n,2k-1}
		\end{array}\right).
	\end{equation} 
Like in the FS-CB case, we compute the exact solution for a particular set of parameters then determine the long-time asymptotics in the reference frame corresponding to each wave. Figure \ref{fig:CB-FB snapshots} shows snapshots of the FB-CB solution for the following parameters:
    \begin{gather}\label{CB-FB par}
        p_1  = \frac{1}{5}+ \frac{i\pi}{4},\quad p_2  = \frac{2}{5},\quad 
		\bd C_1 =  \left(\begin{array}{cc}
			3 & 1 \\ 2 & 1
		\end{array}\right), \quad \bd C_2 =  \left(\begin{array}{cc}
			1 & 1/2 \\ 1 & 1/2
		\end{array}\right).
    \end{gather}
\paragraph{Long-time asymptotics of the FB-CB solutions.}
	We express the solution $\bd q_n(t)$ for the same set of parameters as in Eq. \eqref{CB-FB par} in terms of the real and imaginary part of $E_i$ and $\widetilde{E}_i$ for $i = 1, 2$. In the reference frame of the composite breather, $\xi_1$ is fixed and $\xi_2$ is given in Eq.~\eqref{e:xi2vsxi1}. Assuming as before $v_2>v_1$, we find : 
    \begin{subequations}\label{CB-FB asymptotics ref1}
	\begin{alignat}{2}
		& t\to -\infty \implies e^{\xi_2}\to\infty\colon \qquad && q^{(1)}_n(t) \sim \frac{c_1 e^{\xi_1 + i \nu_1} + c_2e^{\xi_1+ i \widetilde{\nu}_1} + e^{2\xi_1}\left(c_3 e^{\xi_1+ i \widetilde{\nu}_1} + c_4 e^{\xi_1+ i\nu_1}\right)}{s_0 +  e^{2\xi_1} \left[ s_1 + s_2 e^{-i(\nu_1 - \widetilde{\nu}_1)}+ s_3 e^{i(\nu_1 - \widetilde{\nu}_1)}\right] + s_4 e^{4\xi_1} },\\
		&  && q^{(2)}_n(t) \sim \frac{d_1 e^{\xi_1 + i \nu_1} + d_2e^{\xi_1+ i \widetilde{\nu}_1} + e^{2\xi_1}\left(d_3 e^{\xi_1+ i \widetilde{\nu}_1} + d_4 e^{\xi_1+ i\nu_1}\right)}{s_0+  e^{2\xi_1} \left[ s_1 + s_2 e^{-i(\nu_1 - \widetilde{\nu}_1)}+ s_3 e^{i(\nu_1 - \widetilde{\nu}_1)}\right] + s_4 e^{4\xi_1} },\\
		& t\to \infty \implies e^{\xi_2}\to 0\colon\qquad && q^{(1)}_n(t) \sim \frac{\alpha_1 e^{\xi_1 + i \nu_1} + \alpha_2e^{\xi_1+ i \widetilde{\nu}_1} + e^{2\xi_1}\left(\alpha_3 e^{\xi_1+ i \widetilde{\nu}_1} + \alpha_4 e^{\xi_1+ i\nu_1}\right)}{\beta_0 +  e^{2\xi_1} \left[ \beta_1 + \beta_2 e^{-i(\nu_1 - \widetilde{\nu}_1)}+ \beta_3 e^{i(\nu_1 - \widetilde{\nu}_1)}\right] + \beta_4 e^{4\xi_1} },\\
		&  && q^{(2)}_n(t) \sim \frac{\lambda_1 e^{\xi_1 + i \nu_1} + \lambda_2 e^{\xi_1+ i \widetilde{\nu}_1} + e^{2\xi_1}\left(\lambda_3 e^{\xi_1+ i \widetilde{\nu}_1} + \lambda_4 e^{\xi_1+ i\nu_1}\right)}{\beta_0 +  e^{2\xi_1} \left[ \beta_1 + \beta_2 e^{-i(\nu_1 - \widetilde{\nu}_1)}+ \beta_3 e^{i(\nu_1 - \widetilde{\nu}_1)}\right] + \beta_4 e^{4\xi_1} },
	\end{alignat}
    \end{subequations}
	where the values for the coefficients of $q^{(1)}_n(t)$ and $q^{(2)}_n(t)$ in Eqs. \eqref{CB-FB asymptotics ref1} are given in Appendix \ref{ssec: FB-CB coefs}. Similarly to the FS-CB case, and consistent with the prediction in \cite{13} via the Manakov method, Eq.~\eqref{CB-FB asymptotics ref1} confirms that the CB also remains a CB after interacting with the FB. 
	
    Similarly, in the reference frame of the FB, $\xi_2$ is fixed, and $\xi_1$ is expressed via Eq.\eqref{e:xi1vsxi2}. We find
    \begin{subequations}\label{CB-FB asymptotics ref2}
    \begin{gather}
		 t\to -\infty \implies e^{\xi_1}\to 0\colon \quad q^{(1)}_n(t) \sim \frac{m_1 e^{\xi_2+ i\nu_2} + m_2 e^{\xi_2+ i\widetilde{\nu}_2}}{\rho_1 + w_1e^{\xi_2} },\quad  q^{(2)}_n(t)\sim\frac{\sigma_1 e^{\xi_2+ i\nu_2} + \sigma_2e^{\xi_2+ i\widetilde{\nu}_2}}{\rho_1 + w_1e^{2\xi_2} },\\
		t\to \infty \implies e^{\xi_1}\to \infty\colon \quad q^{(1)}_n(t)\sim \frac{\widehat{m}_1 e^{\xi_2+i\nu_2} + \widehat{m}_2e^{\xi_2+i\widetilde{\nu}_2}}{\widehat{\rho}_0 + \widehat{\rho}_1 e^{2\xi_2} },\quad q^{(2)}_n(t)\sim \frac{\widehat{\sigma}_1 e^{\xi_2+i\nu_2} + \widehat{\sigma}_2e^{\xi_2+ i\widetilde{\nu}_2}}{\widehat{\rho}_0 + \widehat{\rho}_1 e^{2\xi_2}},
	\end{gather}
    \end{subequations}
    where the values of the coefficients of $q^{(1)}_n(t)$ and $q^{(2)}_n(t)$ in the direction of  $t\to \pm \infty$ are given in Appendix \ref{ssec: FB-CB coefs}. Comparing the form of the asymptotics of $q^{(1)}_n(t)$ and $q^{(2)}_n(t)$ to the 1-soliton solution in Section \ref{ssec:FB1} and Section \ref{ssec:CB1}, Eq.~\eqref{CB-FB asymptotics ref2} confirms the behaviour observed in Fig. \ref{fig:CB-FB snapshots}: the FB emerges as a FB after interacting with the CB. The above asymptotic expansions have been validated numerically, the plots are omitted for brevity.

	\subsection{Composite breather - composite breather (CB-CB)}
\begin{figure}[H]
		\centering
		\includegraphics[width=0.75\linewidth]{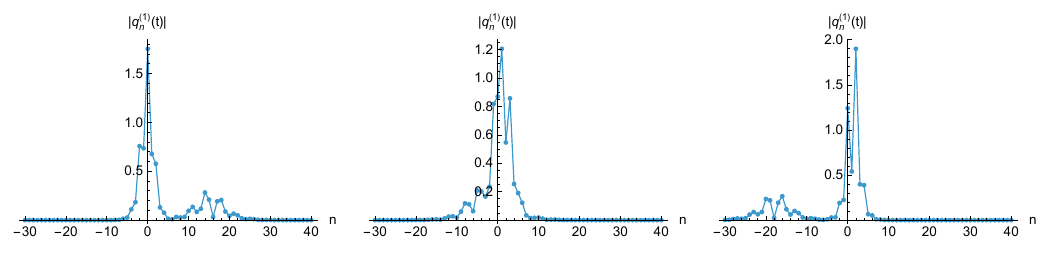}
		\includegraphics[width=0.75\linewidth]{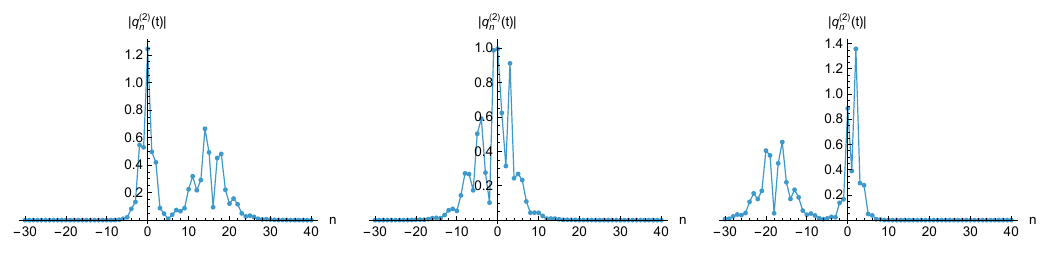}
		\caption{Snapshots at times of a CB-CB solution with parameters as in Eq. \eqref{CB-CB par}.}
		\label{fig:CB-CB snapshots}
	\end{figure}

	The corresponding norming constant matrices are given as
	\begin{align}
		\bd C_1 =  \left(\begin{array}{cc}
			\gamma_1 & \epsilon_1 \\ \gamma_2 & \epsilon_2
		\end{array}\right), \quad \bd C_2 =  \left(\begin{array}{cc}
			\kappa_1 & \delta_1 \\ \kappa_2 & \delta_2
		\end{array}\right),\quad  \epsilon_j, \gamma_j, \kappa_j, \delta_j \in \C, \quad j \in \{1,2\}.
	\end{align}
We solve for { $g_{n,2k-1}, h_{n,2k-1}$ and $f_{n,2k}$} order by order in $\varepsilon$ in the perturbation-like expansion starting with the ansatz:
	\begin{align*}
		g_{n,1} &= \gamma_1^{*} E_1 + \epsilon_1\widetilde{E}_1 + \kappa_1^{*}E_2 + \delta_1\widetilde{E}_2,\\
		h_{n,1} &= \gamma_2^{*} E_1 + \epsilon_2\widetilde{E}_1 + \kappa_2^{*}E_2 + \delta_2\widetilde{E}_2,
	\end{align*}
and obtain the following expression for the CB-CB solution:
	\begin{equation}
		\bd q_n(t) = \left(\begin{array}{c}
			q^{(1)}_n(t)\\ q^{(2)}_n(t)\end{array}\right) = \frac{1}{1+ \sum_{k=1}^{4}f_{n,2k}}\left(\begin{array}{c}
			\sum_{k=1}^{4}g_{n,2k-1}\\ \sum_{k=1}^{4} h_{n,2k-1}
		\end{array}\right).
	\end{equation} 
In Fig.~\ref{fig:CB-CB snapshots} snapshots of a CB-CB solution are provided for the following parameter choices:
    \begin{gather}\label{CB-CB par}
        p_1  = \frac{2}{5}+  \frac{i\pi}{4},\quad  p_2 = 1, \quad 
		\bd C_1 =  \left(\begin{array}{cc}
			2 & 1 \\ 5 & 2
		\end{array}\right),\quad  \bd C_2 =  \left(\begin{array}{cc}
			3 & 4 \\ 2 & 3
		\end{array}\right).
    \end{gather}

\paragraph{Long-time asymptotics of the CB-CB solutions.} Similarly to previous cases, the solution $\bd q_n(t)$ is expressed in terms of the real and imaginary parts of the exponential functions $E_i$ and $\widetilde{E}_i$ for $i= 1, 2$ associating with the CB ``1" and CB ``2" respectively. In the reference frame of the CB ``1", we fix $\xi_1$, and express $\xi_2$ as in Eq. \eqref{e:xi2vsxi1}. Assuming $v_2>v_1$, we find
	\begin{subequations}\label{CB-CB asymptotics ref1}
    \begin{gather}
		t\to -\infty \implies e^{\xi_2} \to \infty \colon \quad  q^{(1)}_n(t) \sim \frac { c_1 e^{\xi_1 + i\nu_1} + c_2 e^{\xi_1 + i\widetilde{\nu}_1} + e^{2\xi_1}\left( c_3 e^{\xi_1 +i \nu_1} + c_4 e^{\xi_1 + i \widetilde{\nu_1}}\right)}{r_0  + e^{2\xi_1}\left(r_1+  r_2 e^{i (\nu_1 - \widetilde{\nu}_1)}+ r_3 e^{-i(\nu_1 - \widetilde{\nu}_1)} \right)+ r_4 e^{4\xi_1} },\\        
		\qquad \qquad\qquad\qquad\qquad\qquad  q^{(2)}_n(t) \sim \frac{d_1 e^{\xi_1+ \nu_1} + d_2 e^{\xi_1 + i\widetilde{\nu}_1} +e^{2\xi_1}\left( d_3 e^{\xi_1 +i \nu_1} + d_4 e^{\xi_1 + i \widetilde{\nu_1}}\right)}{r_0  + e^{2\xi_1}\left(r_1 + r_2 e^{i (\nu_1 - \widetilde{\nu}_1)}+ r_3 e^{-i(\nu_1 - \widetilde{\nu}_1)} \right)+ r_4 e^{4\xi_1} },\\
		t\to \infty \implies e^{\xi_2} \to 0 \colon \quad  q^{(1)}_n(t) \sim \frac{ \hat{c}_1 e^{\xi_1 + i \nu_1} + \hat{c}_2e^{\xi_1 +i \widetilde{\nu}_1} +  e^{2\xi_1}\left(\hat{c}_3 e^{\xi_1+ i\nu_1} + \hat{c}_4 e^{\xi_1+ i \widetilde{\nu}_1}\right)}{\hat{r}_0  + e^{2\xi_1}\left( \hat{r}_1+  \hat{r}_2 e^{i (\nu_1 - \widetilde{\nu}_1)}+ \hat{r}_3 e^{-i(\nu_1 - \widetilde{\nu}_1)} \right) +\hat{r}_4 e^{4\xi_1}},\\
		\qquad \qquad\qquad\qquad\qquad\quad q^{(2)}_n(t) \sim \frac{ \hat{d}_1 e^{\xi_1 + i\nu_1} + \hat{d}_2 e^{\xi_1 + i\widetilde{\nu}_1} +  e^{2\xi_1}\left(\hat{d}_3 e^{\xi_1+ i\nu_1} + \hat{d}_4 e^{\xi_1+ i \widetilde{\nu}_1}\right)}{\hat{r}_0  + e^{2\xi_1}\left( \hat{r}_1+ \hat{r}_2 e^{i (\nu_1 - \widetilde{\nu}_1)}+ \hat{r}_3 e^{-i(\nu_1 - \widetilde{\nu}_1)} \right) + \hat{r}_4 e^{4\xi_1}},
	\end{gather}
    \end{subequations}
where the values of the coefficients of the above asymptotics expressions of $q^{(1)}_n(t)$ and $q^{(2)}_n(t)$ are given in Eq. \eqref{CB-CB coefs ref1} in Appendix \ref{ssec: CB-CB coefs}. 
	
In the reference frame of the CB ``2", we fix $\xi_2$, recast $\xi_1$ according to Eq. \eqref{e:xi1vsxi2} and observe a similar asymptotic behaviour for $\bd q_n(t)$: 
    \begin{subequations}\label{CB-CB asymptotics ref2}
    \begin{gather}
		t\to -\infty \implies e^{\xi_1}\to 0 \colon \quad  q^{(1)}_n(t) \sim \frac { \mu_1 e^{\xi_2 + i\nu_2} + \mu_2 e^{\xi_2 + i\widetilde{\nu}_2} + e^{2\xi_2}\left( \mu_3 e^{\xi_2 +i \nu_2} + \mu_4 e^{\xi_2 + i \widetilde{\nu_2}}\right)}{\chi_0  + e^{2\xi_2}\left(\chi_1+  \chi_2 e^{i (\nu_2 - \widetilde{\nu}_2)}+ \chi_3 e^{-i(\nu_2 - \widetilde{\nu}_2)} \right)+ \chi_4 e^{4\xi_2} },\\        
		\qquad \qquad \qquad\qquad \qquad \quad  q^{(2)}_n(t) \sim \frac{\sigma_1 e^{\xi_2+ \nu_2} + \sigma_2 e^{\xi_2 + i\widetilde{\nu}_2} +e^{2\xi_2}\left( \sigma_3 e^{\xi_2 +i \nu_2} + \sigma_4 e^{\xi_2 + i \widetilde{\nu}_2}\right)}{\chi_0  + e^{2\xi_2}\left(\chi_1 + \chi_2 e^{i (\nu_2 - \widetilde{\nu}_2)}+ \chi_3 e^{-i(\nu_2 - \widetilde{\nu}_2)} \right)+ \chi_4 e^{4\xi_2} }.\\
		t\to \infty \implies e^{\xi_1}\to \infty\colon \quad q^{(1)}_n(t) \sim \frac{ \hat{\mu}_1 e^{\xi_2 + i \nu_2} + \hat{\mu}_2e^{\xi_2 +i \widetilde{\nu}_2} +  e^{2\xi_2}\left(\hat{\mu}_3 e^{\xi_2+ i\nu_2} + \hat{\mu}_4 e^{\xi_2+ i \widetilde{\nu}_2}\right)}{\hat{\chi}_0  + e^{2\xi_2}\left( \hat{\chi}_1+  \hat{\chi}_2 e^{i (\nu_2 - \widetilde{\nu}_2)}+ \hat{\chi}_3 e^{-i(\nu_2 - \widetilde{\nu}_2)} \right) + \hat{\chi}_4 e^{4\xi_2}},\\
		\qquad \qquad \qquad\qquad \qquad\quad q^{(2)}_n(t) \sim \frac{ \hat{\sigma}_1 e^{\xi_2 + i\nu_2} + \hat{\sigma}_2 e^{\xi_2 + i\widetilde{\nu}_2} +  e^{2\xi_2}\left(\hat{\sigma}_3 e^{\xi_2 + i\nu_2} + \hat{\sigma}_4 e^{\xi_2+ i \widetilde{\nu}_2}\right)}{\hat{\chi}_0  + e^{2\xi_2}\left( \hat{\chi}_1+ \hat{\chi}_2 e^{i (\nu_2 - \widetilde{\nu}_2)}+ \hat{\chi}_3 e^{-i(\nu_2 - \widetilde{\nu}_2)} \right) + \hat{\chi}_4 e^{4\xi_2}},
	\end{gather}
    \end{subequations}
where the values of the coefficients of the asymptotics expressions for $q^{(1)}_n(t)$ and $q^{(2)}_n(t)$ above are given in Eqs \eqref{CB-CB coefs ref2} in  Appendix \ref{ssec: CB-CB coefs}. We note that since the asymptotic expressions in Eqs. \eqref{CB-CB asymptotics ref1} and \eqref{CB-CB asymptotics ref2} have similar form to the 1-soliton {CB} solution in Section. \ref{ssec:CB1}, we conclude that a CB retains its nature after interacting with another CB. As in all previous cases, the above asymptotic expansions have been validated numerically but we omit the plots for brevity.

\section{Conclusions}

In this work, we applied Hirota’s bilinear formalism to the focusing IDM system and systematically constructed explicit solutions describing fundamental solitons, fundamental breathers, composite breathers, and their interactions. Building upon the bilinear representation of the system, we showed that Hirota’s method provides a direct and efficient framework for deriving these coherent structures in explicit closed form, as finite sums of exponentials, thereby avoiding the technical complexity associated with determinant representations arising in the IST approach. In particular, the construction of fundamental and composite breathers through Hirota’s method is new, and extends previous bilinear studies focused primarily on fundamental soliton solutions.

A central outcome of this work is the transparent characterization of interaction dynamics among the various coherent structures supported by the IDM system. The explicit bilinear formulas make it possible to compute the long-time asymptotics directly in both time directions, yielding precise information on phase shifts, polarization changes, and interaction-induced effects. In contrast with determinant-based formulas, whose asymptotic analysis often requires delicate handling of degenerate leading-order contributions, the Hirota representation naturally separates dominant exponential contributions in different asymptotic regimes. This substantially simplifies both the analytical study and the numerical visualization of multi-soliton and multi-breather interactions. 

An important direction for future work concerns the stability theory of the coherent structures constructed here. In the continuous Manakov system, recent results \cite{Pel2026} established the spectral stability of nondegenerate vector solitons and the orbital stability of breather solutions for coupled NLS equations. In particular, nondegenerate vector solitons were shown to remain spectrally stable even though the associated linearized operator possesses embedded or isolated eigenvalues of negative Krein signature, while the orbital stability of breathers was proved through a Lyapunov functional approach combined with the squared-eigenfunction formalism arising from integrability. The stability of the discrete breathers of the IDM system is a completely open problem,  and extending these results to the IDM system represents a highly nontrivial and mathematically relevant task. We expect that the interplay between discreteness, vectorial interactions, and breather dynamics may also reveal new stability phenomena absent in the continuous setting. For instance, the fact that a fundamental soliton generically turns into a fundamental breather (i.e., it acquires a nontrivial projection in the orthogonal polarization) upon interacting with a fundamental breather, suggests the latter to be the stable coherent structure in the discrete setting, in spite of the fact that: (i) fundamental solitons are the natural discretization of a bright soliton of the Manakov system, and (ii) fundamental and composite breathers are purely discrete solutions, and have no analog in the continuous limit.

The present work also highlights the broader potential of Hirota’s method for the study of coherent structures on nonzero backgrounds. There is a vast literature on discrete solitons, breathers and rogue wave solutions of the AL lattice on a nonzero background, both in the focusing and defocusing dispersion regimes, and several recent works have sparked renewed interest in these systems \cite{KV,MO06,ABP07,OY14,vdM15,PV16,P16,OP19,CS24,WM25,MRC26}, including the construction of solitons and breathers sitting on periodic or quasi-periodic wave backgrounds \cite{Pel24,Bar26}.
Some of the discrete coherent structures of the scalar AL lattice on a nonzero background have been generalized to the multicomponent setting \cite{Ank13,FO17}, but many more { are} still to be discovered. 
At present, such configurations remain largely inaccessible within the standard IST framework, whereas Hirota’s method provides a direct constructive avenue for their analysis. In particular, bilinear techniques have played a fundamental role in obtaining higher-order rational solutions and rogue-wave-type excitations in many continuous and discrete integrable models. The present bilinear formulation therefore provides a natural starting point for the systematic derivation of vector discrete breathers and rogue waves, their interaction properties, and their dynamics over nontrivial backgrounds. 


\section*{Acknowledgements}
BP gratefully acknowledges partial support for this work from the NSF, under grant DMS-2406626, the Fulbright Foundation in Greece and the Fulbright program, and the Mathematics Department of the University of Ioannina, Greece, for the kind hospitality during the completion of this work. { UL and AC thank Willy Hereman for valuable discussions on the subject of this work. UL thanks Dr. Devanayagam Palaniappan for his support, and gratefully acknowledges postdoctoral funding from Texas A\&M University - Corpus Christi. }

\section*{Conflict of interest}
The authors declare that they have no conflict of interest.

\appendix

\section{Appendix}

\subsection{Coefficients of the FB-FB solution}\label{ssec: FB-FB coefs}
We give below the coefficients of the exponentials \eqref{e:FBFBexp} of the FB-FB solution \eqref{e:FBFB}:
\begin{gather}
	\bd \gamma = \left(\begin{array}{c}
		\gamma_1 \\ \gamma_2 
		\end{array}\right),\quad  \bd \gamma^{\perp} = \left(\begin{array}{c}
			-\gamma_2 \\ \gamma_1 
	\end{array}\right),\quad \bd \kappa = \left(\begin{array}{c}
		\kappa_1 \\ \kappa_2 \end{array}\right), \quad  \bd \kappa^{\perp} = \left(\begin{array}{c}
		-\kappa_2 \\ \kappa_1 \end{array}\right),
\end{gather}
\begin{subequations}
\begin{alignat}{3}
		\alpha_1 &= \langle \mu_1^*\bd\gamma^*, \mu_1 \bd\gamma \rangle A_{(1,1^*)},\quad \alpha_2 =  \langle \mu_1^*\bd\gamma^*, \omega_1\bd \kappa \rangle A_{1, 2^{*}},\quad  &\alpha_3 =  \langle \mu_1^{*} \bd \gamma^{*}, \omega_2 \bd \kappa^{\perp}\rangle A_{(1,\widetilde{2}^*)},\\
		\alpha_4 &=  \langle \omega_1^*\bd \kappa^*, \mu_1\bd \gamma \rangle A_{(2,1^*)},\quad
		\alpha_5 = \langle \omega_1^*\bd \kappa^*, \omega_1, \bd \kappa \rangle A_{(2,2^*)},\quad &\alpha_6 =  \langle \omega_1^*\kappa^*, \mu_2^* \bd \gamma^{\perp,*}  \rangle A_{(2, \widetilde{1}^*)},\\
		\alpha_7 &=  \langle \mu_2 \bd \gamma^{\perp}, \omega_1 \bd \kappa \rangle A_{(\widetilde{1},2^*)},\quad \alpha_8 =  \langle \mu_1^* \bd \gamma^*, \mu_2^* \bd \gamma^{\perp, *}\rangle A_{(1, \widetilde{1}^*)},    \quad &\alpha_9 =  \langle \mu_2 \bd \gamma^{\perp}, \omega_2, \bd \kappa^{\perp}\rangle A_{(\widetilde{1}, \widetilde{2}^*)} ,\\
		\alpha_{10} &= \langle \omega_2, \bd\kappa^{\perp}, \mu_1, \bd \gamma\rangle A_{(\widetilde{2}, 1^{*})} , \quad \alpha_{11} =  \langle \omega_2 \bd \kappa^{\perp}, \mu_2^* \bd \gamma^{\perp, *}  \rangle A_{(\widetilde{2}, \widetilde{1}^*)} ,\quad &\alpha_{12} =  \langle \omega_2 \bd \kappa^{\perp},  \omega_2^* \bd \kappa^{\perp,*}\rangle A_{(\widetilde{2}, \widetilde{2}^*)} ,
		\end{alignat}
\end{subequations}
\begin{subequations}
\begin{align}	
		\beta_1 &= P_{(1,2)}P_{(1,1^*)}\alpha_4\mu_1^*\gamma_1^* - P_{(1,2)}P_{(2,1^*)}\alpha_1\omega_1^*\kappa_1^*, \\
		\beta_2 &= P_{(1,2)}P_{(1,2^*)}\alpha_5\mu_1^*\gamma_1^* - P_{(1,2)}P_{(2,2^*)}\alpha_2\omega_1^*\kappa_1^* ,\\
		\beta_3 &= P_{(1,\widetilde{1})}P_{(1,2^*)}\alpha_7\mu_1^*\gamma_1^* - P_{(1,\widetilde{1})}P_{(\widetilde{1},2^*)}\alpha_2\mu_2(-\gamma_2),\\
		\beta_4 &=P_{(1,\widetilde{1})}P_{(1,\widetilde{2}^*)}\alpha_9\mu_1^*\gamma_1^* - P_{(1,\widetilde{1})}P_{(\widetilde{1},\widetilde{2}^*)}\alpha_3\mu_2(-\gamma_2),\\
		\beta_5 &= P_{(2,\widetilde{1})}P_{(2,2^*)}\alpha_7\omega_1^*\kappa_1^* - P_{(2,\widetilde{1})}P_{(\widetilde{1},2^*)}\alpha_5\mu_2(-\gamma_2),\\
		\beta_6 &=P_{(2,\widetilde{1})}P_{(2,\widetilde{1}^*)}\alpha_8\omega_1^*\kappa_1^* - P_{(2,\widetilde{1})}P_{(\widetilde{1},\widetilde{1}^*)}\alpha_6\mu_2(-\gamma_2),\\
		\beta_7 &= P_{(1,\widetilde{2})}P_{(1,1^*)}\alpha_{10}\mu_1^*\gamma_1^* - P_{(1,\widetilde{2})}P_{(\widetilde{2},1^*)}\alpha_1\omega_2(-\kappa_2) ,\\
		\beta_8 &= P_{(1,\widetilde{2})}P_{(1,\widetilde{2}^*)}\alpha_{12}\mu_1^*\gamma_1^* - P_{(1,\widetilde{2})}P_{(\widetilde{2},\widetilde{2}^*)}\alpha_3\omega_2(-\kappa_2) ,\\
		\beta_9 &= P_{(2,\widetilde{2})}P_{(2,1^*)}\alpha_{10}\omega_1^*\kappa_1^* - P_{(2,\widetilde{2})}P_{(\widetilde{2},1^*)}\alpha_4\omega_2(-\kappa_2) ,\\
		\beta_{10} &= P_{(2,\widetilde{2})}P_{(2,\widetilde{1}^*)}\alpha_{11}\omega_1^*\kappa_1^* - P_{(2,\widetilde{2})}P_{(\widetilde{2},\widetilde{1}^*)}\alpha_6\omega_2(-\kappa_2) ,\\
		\beta_{11} &=P_{(\widetilde{1},\widetilde{2})}P_{(\widetilde{1},\widetilde{1}^*)}\alpha_{11}\mu_2(-\gamma_2) - P_{(\widetilde{1},\widetilde{2})}P_{(\widetilde{2},\widetilde{1}^*)}\alpha_8\omega_2(-\kappa_2) ,\\
		\beta_{12} &=P_{(\widetilde{1},\widetilde{2})}P_{(\widetilde{1},\widetilde{2}^*)}\alpha_{12}\mu_2(-\gamma_2) - P_{(\widetilde{1},\widetilde{2})}P_{(\widetilde{2},\widetilde{2}^*)}\alpha_9\omega_2(-\kappa_2),
		\end{align}
\end{subequations}        
\begin{subequations}		
\begin{align}		
		\delta_1 &= P_{(1,2)}P_{(1,1^*)}\alpha_4\mu_1^*\gamma_2^* - P_{(1,2)}P_{(2,1^*)}\alpha_1\omega_1^*\kappa_2^* ,\\
		\delta_2 &= P_{(1,2)}P_{(1,2^*)}\alpha_5\mu_1^*\gamma_2^* - P_{(1,2)}P_{(2,2^*)}\alpha_2\omega_1^*\kappa_2^* ,\\
		\delta_3 &= P_{(1,\widetilde{1})}P_{(1,2^*)}\alpha_7\mu_1^*\gamma_2^* - P_{(1,\widetilde{1})}P_{(\widetilde{1},2^*)}\alpha_2\mu_2\gamma_1,\\
		\delta_4 &=P_{(1,\widetilde{1})}P_{(1,\widetilde{2}^*)}\alpha_9\mu_1^*\gamma_2^* - P_{(1,\widetilde{1})}P_{(\widetilde{1},\widetilde{2}^*)}\alpha_3\mu_2\gamma_1,\\
		\delta_5 &= P_{(2,\widetilde{1})}P_{(2,2^*)}\alpha_7\omega_1^*\kappa_2^* - P_{(2,\widetilde{1})}P_{(\widetilde{1},2^*)}\alpha_5\mu_2\gamma_1,\\
		\delta_6 &=P_{(2,\widetilde{1})}P_{(2,\widetilde{1}^*)}\alpha_8\omega_1^*\kappa_2^* - P_{(2,\widetilde{1})}P_{(\widetilde{1},\widetilde{1}^*)}\alpha_6\mu_2\gamma_1,\\
		\delta_7 &= P_{(1,\widetilde{2})}P_{(1,1^*)}\alpha_{10}\mu_1^*\gamma_2^* - P_{(1,\widetilde{2})}P_{(\widetilde{2},1^*)}\alpha_1\omega_2\kappa_1 ,\\
		\delta_8 &= P_{(1,\widetilde{2})}P_{(1,\widetilde{2}^*)}\alpha_{12}\mu_1^*\gamma_2^* - P_{(1,\widetilde{2})}P_{(\widetilde{2},\widetilde{2}^*)}\alpha_3\omega_2\kappa_1 ,\\
		\delta_9 &= P_{(2,\widetilde{2})}P_{(2,1^*)}\alpha_{10}\omega_1^*\kappa_2^* - P_{(2,\widetilde{2})}P_{(\widetilde{2},1^*)}\alpha_4\omega_2\kappa_1 ,\\
		\delta_{10} &= P_{(2,\widetilde{2})}P_{(2,\widetilde{1}^*)}\alpha_{11}\omega_1^*\kappa_2^* - P_{(2,\widetilde{2})}P_{(\widetilde{2},\widetilde{1}^*)}\alpha_6\omega_2\kappa_1 ,\\
		\delta_{11} &=P_{(\widetilde{1},\widetilde{2})}P_{(\widetilde{1},\widetilde{1}^*)}\alpha_{11}\mu_2\gamma_1 - P_{(\widetilde{1},\widetilde{2})}P_{(\widetilde{2},\widetilde{1}^*)}\alpha_8\omega_2\kappa_1 ,\\
		\delta_{12} &=P_{(\widetilde{1},\widetilde{2})}P_{(\widetilde{1},\widetilde{2}^*)}\alpha_{12}\mu_2\gamma_1 - P_{(\widetilde{1},\widetilde{2})}P_{(\widetilde{2},\widetilde{2}^*)}\alpha_9\omega_2\kappa_1,
	\end{align}
\end{subequations}
\begin{subequations}	
\begin{gather}
	\rho_1 = A_{(1,1^*, 2, 2^*)}\left( \mu_1^*\gamma_1^* \beta_2^*+ \mu_1^*\gamma_2^*\delta_2^* + \omega_1^*\kappa_1^* \beta_1^* + \omega_1^*\kappa_2^*\delta_1^* \right) \\ 
    +A_{(1,1^*, 2, 2^*)}\left( \mu_1\gamma_1 \beta_2+ \mu_1\gamma_2\delta_2 + \omega_1\kappa_1 \beta_1 +\omega_1\gamma_2 \delta_1  \right) \nonumber\\
    - S^2_{[(1,1^*),(2,2^*)]}\alpha_{1}\alpha_{5}- S^2_{[(1,2^*),(2,1^*)]}\alpha_{2}\alpha_{4},  \nonumber\\
    \nonumber
	\rho_2 = A_{(2,2^*,3,3^*)}\left[  \omega_1^*\kappa_1^* \beta_6^*+ \omega_1^*\kappa_2^*\delta_6^* + \mu_2^*(-\gamma_2^*) \beta_5^* + \mu_2^*\gamma_2^* \delta_5^* \right] \\
    + A_{(2,2^*,3,3^*)}\left[\omega_1\kappa_1 \beta_6+ \omega_1\kappa_2\delta_6 + \mu_2(-\gamma_2) \beta_5 + \mu_2\gamma_2 \delta_5\right]\nonumber\\
    - S^2_{[(2,2^*),(3,3^*)]}\alpha_{5}\alpha_{8}- S^2_{[(2,3^*),(3,2^*)]}\alpha_{6}\alpha_{7},  \nonumber\\
    \nonumber\\
	\rho_3 = A_{(1,1^*,4,4^*)}\left[\mu_1^*\gamma_1^* \beta_8^*+
 \mu_1^*\gamma_2^*\delta_8^* + 
 \omega_2^*(-\kappa_2^*) \beta_7^* + 
 \omega_2^*\kappa_1^* \delta_7^* \right] \\
 + A_{(1,1^*,4,4^*)} \left[\mu_1\gamma_1 \beta_8+
 \mu_1\gamma_2\delta_8 + 
 \omega_2(-\kappa_2) \beta_7 +
 \omega_2\kappa_1 \delta_7 \right]\nonumber\\
 - S^2_{[(1,1^*),(4,4^*)]}\alpha_{1}\alpha_{12}- S^2_{[(1,4^*),(4,1^*)]}\alpha_{3}\alpha_{10},  \nonumber\\
 \nonumber\\
	\rho_4 = A_{(3,3^*,4,4^*,)}\left[
	\mu_2(-\gamma_2) \beta_{12}^*+
	\mu_2\gamma_1\delta_{12}^* + 
	\omega_2(-\kappa_2) \beta_{11}^* + 
	\omega_2\kappa_1 \delta_{11}^* \right] \\
    + A_{(3,3^*,4,4^*,)}\left[ 
	\mu_2^*(-\gamma_2^*) \beta_{12}+
	\mu_2^*\gamma_1^*\delta_{12} + 
	\omega_2^*(-\kappa_2^*) \beta_{11} + 
	\omega_2^*\kappa_1^* \delta_{11} 
	 \right]\nonumber\\ 
     - S^2_{[(3,3^*),(4,4^*)]}\alpha_{8}\alpha_{12}- S^2_{[(3,4^*),(4,3^*)]}\alpha_{9}\alpha_{11},  \nonumber\\
     \nonumber\\
	\rho_5 =  A_{(1,2^*,\widetilde{1},\widetilde{2}^*)}\left[
	\mu_1^*\gamma_1^* \beta_{10}^*+
	\mu_1^*\gamma_2^*\delta_{10}^* + 
	\mu_2^*(-\gamma_2^*) \beta_9^* + 
	\mu_2^*\gamma_1^* \delta_9^* \right] \\
    + A_{(1,2^*,\widetilde{1},\widetilde{2}^*)} \left[ 
	\omega_1\kappa_1 \beta_4+
	\omega_1\kappa_2\delta_4 + 
	\omega_2^*(-\kappa_2^*) \beta_3 +
	\omega_2^*\kappa_1^* \delta_3 
	\right]\nonumber\\
    - S^2_{[(1,2^*),(\widetilde{1},\widetilde{2}^*)]}\alpha_{2}\alpha_{9}- S^2_{[(1,\widetilde{2}^*),(2^*,\widetilde{1})]}\alpha_{3}\alpha_{11},  \nonumber\\
    \nonumber \\
	\rho_6 =  A_{(1^*,2,\widetilde{1}^*,\widetilde{2})}\left[
	\mu_1\gamma_1 \beta_{10}+
	\mu_1\gamma_2\delta_{10} + 
	\mu_2(-\gamma_2) \beta_9 + 
	\mu_2\gamma_1 \delta_9 \right] \\
    + A_{(1^*,2,\widetilde{1}^*,\widetilde{2})}\left[ 
	\omega_1^*\kappa_1^* \beta_4^*+
	\omega_1^*\kappa_2^*\delta_4^* + 
	\omega_2(-\kappa_2) \beta_3^* +
	\omega_2\kappa_1 \delta_3^* 
	\right]\nonumber\\
    - S^2_{[(2,\widetilde{1}^*),(1^*, \widetilde{2})]}\alpha_{6}\alpha_{10}- S^2_{[(1^*,2),(\widetilde{1}^*,\widetilde{2})]}\alpha_{4}\alpha_{11},	  \nonumber
	\end{gather}
\end{subequations}
and
\begin{subequations}
	\begin{gather}
	A_{(j,k^*)} = \frac{e^{p_j+p_k^{*}} \Delta h^2}{\left(e^{p_j+p_k^{*}}-1\right)^2},\quad 
	P_{(j,k^*)} = \frac{e^{p_j} + e^{p_k^{*}}}{e^{p_j+p_k^{*}}-1},\quad P_{(j,k)} = \frac{e^{p_j} - e^{p_k}}{e^{p_j+p_k}+1},\quad j,k \in \{1,2\},\\
	S_{[(j,k^*),(l,m^*)]}= \frac{e^{p_j+ p_k^*}- e^{p_l+ p_m^*}}{e^{p_j+p_k^*+p_l+p_m^*}-1}, \quad j,k,l,m \in \{1,2\}.
	\end{gather}
\end{subequations}

\subsection{Coefficients of the FS-CB solution}\label{ssec: FS-CB coefs}
We give below the numerical values of the coefficients of the exponentials for the FS-CB solution in Fig.~\ref{fig:CB-FS snapshots} corresponding to parameters \eqref{FS-CB_par}:
\begin{subequations}
\begin{gather}
 		\gamma_1 = 1, \quad \epsilon_2 = -1/10,\quad \kappa_1 = 1,\\ 
 		\beta_1 = 0.0828 - 0.0736i ,\quad \beta_2 = 1.3277 + 3.5020 i, \quad \beta_3 = -0.0083 - 0.0074i ,\\ \beta_4= -0.0027 - 0.0046 i , \quad \beta_5  =0.1547 - 0.0108 i, \quad \beta_6 =-0.591 - 1.5474 i, \\\beta_7 = -0.0735 - 0.0696i, \quad \beta_8 =0.0115 + 0.0144i,\quad \beta_9= -0.1538 - 0.1021i,\\
		\mu_1 = 0.0058 - 0.071i ,\quad \mu_2 = -0.0011 + 0.0046i,\quad \mu_3 = -0.0079 - 0.0011i,\\
        \nonumber\\
		\gamma_2 = 2, \quad \epsilon_1 = 1/10,\quad  \kappa_2 = 1,\\
		\delta_1 = 0.1657 - 0.1471i ,\quad \delta_2 = -0.2191 + 3.3938i, \quad \delta_3 = 0.0083 + 0.0074i ,\\ \delta_4= -0.0027 - 0.0046i  , \quad \delta_5  = -0.1547 + 0.0108i,\quad \delta_6 = 0.9468 - 2.5685i ,\\\delta_7 = -0.0735 - 0.0696i , \quad \delta_8 = 0.0115 + 0.0144i,\quad \delta_9= 0.1538 + 0.1021i,\\
		\sigma_1 = 0.0843 - 0.0820i ,\quad \sigma_2 = -0.0011 + 0.0046i ,\quad \sigma_3 = 0.0078 + 0.0011i ,\\
		\nonumber\\
		\alpha_1 = 7.4088 ,\quad  \alpha_2 = -0.0280 - 0.0248i , \quad \alpha_3= -0.2050 + 2.3709i ,\quad \alpha_4 = -0.0280 + 0.0248i ,\\
        \alpha_5 = 0.0296 ,\quad \alpha_6 = -0.2050 - 2.3709i ,\quad \alpha_7 = 2.9635,\\
        \nonumber\\
		\rho_1 = 5.1185 ,\quad \rho_2 = 0.0308 - 0.0219i , \quad \rho_3= -0.0388 + 0.0185i  ,\quad  \rho_4 = 0.0308 + 0.0219i ,\\
        \rho_5 = 0.0122748 ,\quad \rho_6 = -0.0388 - 0.0185i  ,\quad \rho_7 = 0.0340743, \quad
		\rho_8 = 0.0055.
 	\end{gather}
\end{subequations}
\subsection{Coefficients of the FB-CB solution}\label{ssec: FB-CB coefs}
We give below the numerical values of the coefficients for the asymptotics expression of the FB-CB solution in Eq. \eqref{CB-FB asymptotics ref1} corresponding to parameters in Eq. \eqref{CB-FB par}:
\begin{gather*}
		c_1 = 1.298 + 7.492 i, \quad  c_2 = 5.328 + 0.503 i, \quad c_3 = 42.966 + 80.38 i , \quad c_4 = 118.571 + 52.037 i ,\\
		d_1 = -0.317 + 3.941 i, \quad d_2 = 0.289 - 0.671 i, \quad  d_3 = 12.237 - 2.205 i , \quad  d_4 = 52.972 + 41.573 i ,\\
		s_0 = 3.704 , \quad  s_1 = 170.84, \quad  s_2 = -2.825 + 4.264i , \quad  s_3 = -2.825 - 4.264 i , \quad s_4 = 1074.44,
	\end{gather*}
    and the coefficients for the $t\to \infty$ directions are
    \begin{gather*}
		\alpha_1= 3 , \quad \alpha_2 = 1, \quad \alpha_3 = 79.155 + 32.513 i , \quad \alpha_4 = 26.385 - 10.838i ,\\
        \lambda_1 = 2, \quad \lambda_2 = 1, \quad \lambda_3 = 52.770 + 21.675 i ,  \quad \lambda_4 = 26.385 - 10.838 i ,\\
        \beta_0= 1, \quad \beta_1 = 92.51, \quad \beta_2 = -2.139 - 0.879 i, \quad \beta_3 = -2.139 + 0.879 i , \quad \beta_4 = 813.626.
	\end{gather*}
The numerical values of the coefficients in the asymptotics expressions of the FB-CB solution in Eq. \eqref{CB-FB asymptotics ref2} corresponding to parameters Eq. \eqref{CB-FB par}:
    \begin{gather*}
		m_1 = 1,\quad  m_2 = -1/2,\quad \sigma_1 = 1,\quad \sigma_2 = 1/2,\quad
		\rho_1 = 1 ,\quad  w_1 = 3.7043,
	\end{gather*}
    and the coefficients in the $t\to \infty$ direction are
    \begin{gather*}
		\widehat{m}_1 = 297.669 - 383.901 i ,\quad  \widehat{m}_2 = -148.834 - 191.95 i ,\\
        \widehat{\sigma}_1 = 297.669 - 383.901 i ,\quad \widehat{\sigma}_2 = 148.834 + 191.95 i ,\\
		\hat{\rho}_0 = 813.626 ,\quad \hat{\rho}_1 = 1074.441 .
	\end{gather*}
\subsection{Coefficients of the CB-CB solution}\label{ssec: CB-CB coefs}
We give the numerical values of the coefficients of the asymptotics expressions for the CB-CB solution in the reference frame of the CB ``1" in Eqs. \eqref{CB-CB asymptotics ref1} corresponding to the parameters in Eq. \eqref{CB-CB par}. 
    \begin{subequations}\label{CB-CB coefs ref1}
    \begin{gather}
		c_1= -0.6249 + 0.53345 i , \quad c_2 = -0.3125 - 0.26672 i, \\ c_3 = -0.0362 + 0.2604 i,\quad 
        c_4= -0.0723 - 0.5208 i,\\
        d_1= -1.5624 + 1.3336 i , \quad d_2 = -0.6249 - 0.5334 i , \\ d_3 = -0.0723 + 0.5208 i , \quad
        d_4=  -0.1809 - 1.3021 i,\\
        r_0= 1.8725 , \quad r_1 = 4.5411 , \quad r_2 = 0.2177 + 0.3408 i , \quad r_3= 0.2177 - 0.3408 i,\quad
        r_4= 0.7669,\\
        \hat{c}_1 =  2, \quad \hat{c}_2 = 1 , \quad \hat{c}_3 = 9.9407 - 8.8284 i , \quad \hat{c}_4 = 19.8813 + 17.6567 i ,\\
        \hat{d}_1 = 5, \quad  \hat{d}_2 = 2, \quad  \hat{d}_3 = 19.8813 - 17.6567 i, \quad  \hat{d}_4 = 49.7034 + 44.1419 i,\\
        \hat{r}_0 = 1, \quad \hat{r}_1 = 50.3801 , \quad \hat{r}_2 = -3.35433 + 2.979 i , \quad \hat{r}_3 = -3.35433 - 2.979 i,\quad 
        \hat{r}_4 = 176.757 .
	\end{gather}
    \end{subequations}
The numerical values of the coefficients of the asymptotics expressions for the CB-CB solution in the reference frame of the CB ``2" in Eqs. \eqref{CB-CB asymptotics ref2} corresponding to the parameters in Eq. \eqref{CB-CB par} are as follows: 
    \begin{subequations}\label{CB-CB coefs ref2}
    \begin{gather}
        \mu_1= 3 , \quad \mu_2 = 2, \quad \mu_3 = 2.7368 ,\quad \mu_4= 4.10518 ,\\
        \sigma_1= 4, \quad \sigma_2 = 3 , \quad \sigma_3 = 4.10518 ,\quad \sigma_4= 5.47357 ,\\
        \chi_0= 1 , \quad \chi_1 = 6.8786 , \quad \chi_2 = -1.8899 , \quad \chi_3= -1.8899 , \quad
        \chi_4= 1.8725,\\
        \hat{\mu}_1 = 66.297 - 95.6042 i , \quad \hat{\mu}_2 = 44.198 + 63.7361 i, \\ \hat{\mu}_3 = 2.9113 - 4.1983 i, \quad \hat{\mu}_4 = 4.367 + 6.2975 i ,\\
        \hat{\sigma}_1 = 88.3961 - 127.472 i, \quad \hat{\sigma}_2 = 66.297 + 95.6042 i , \\  \hat{\sigma}_3 = 4.367 - 6.2975 i, \quad \hat{\sigma}_4 = 5.823 + 8.397i ,\\
        \hat{\chi}_0 = 176.757 , \quad \hat{\chi}_1 = 58.5265 , \quad \hat{\chi}_2 = 5.6369 + 15.0597 i , \\ \hat{\chi}_3 = 5.6369 - 15.0597 i,\quad 
        \hat{\chi}_4 = 0.766923 .
	\end{gather}
    \end{subequations}

\end{document}